\documentclass[journal]{IEEEtran}
\usepackage[T1]{fontenc}
\usepackage[latin9]{inputenc}
\usepackage{color}
\usepackage{float}
\usepackage{amsmath}
\usepackage{amssymb}
\usepackage{graphicx}
\usepackage[bookmarks=true,bookmarksnumbered=true,bookmarksopen=true,bookmarksopenlevel=1,
 breaklinks=false,pdfborder={0 0 0},pdfborderstyle={},backref=false,colorlinks=false]
 {hyperref}
\hypersetup{pdftitle={Your Title},
 pdfauthor={Your Name},
 pdfpagelayout=OneColumn, pdfnewwindow=true, pdfstartview=XYZ, plainpages=false}

\makeatletter

\floatstyle{ruled}
\newfloat{algorithm}{tbp}{loa}
\providecommand{\algorithmname}{Algorithm}
\floatname{algorithm}{\protect\algorithmname}

\usepackage[caption=false,font=footnotesize]{subfig}

\makeatother

\begin{document}
\title{Angular Sensing by Highly Reconfigurable Pixel Antennas with Joint
Radiating Aperture and Feeding Ports Reconfiguration}
\author{Zixiang Han,~\IEEEmembership{Member,~IEEE,}~Hanning Wang,~Shiwen
Tang,~\IEEEmembership{Member,~IEEE,}~and~Yujie Zhang,~\IEEEmembership{Member,~IEEE}\thanks{This work was supported in part by NTU Start-up Grant and in part
by the Ministry of Education (Singapore) under AcRF TIER1 (RS30/25).
\textit{(Corresponding author: Yujie Zhang, }e-mail: yujie.zhang@ntu.edu.sg\textit{)}}\thanks{Z. Han and H. Wang are with the Future Research Lab, China Mobile
Research Institute, 100053, Beijing, China.}\thanks{S. Tang is with the Department of Electrical and Computer Engineering,
National University of Singapore, Singapore.}\thanks{Y. Zhang is with School of Electrical and Electronic Engineering,
Nanyang Technological University, Singapore.}}
\maketitle
\begin{abstract}
Angular sensing capability is realized using highly reconfigurable
pixel antenna (HRPA) with joint radiating aperture and feeding ports
reconfiguration. Pixel antennas represent a general class of reconfigurable
antenna designs in which the radiating surface, regardless of its
shape or size, is divided into sub-wavelength elements called pixels.
Each pixel is connected to its neighboring elements through radio
frequency switches. By controlling pixel connections, the pixel antenna
topology can be flexibly adjusted so that the resulting radiation
pattern can be reconfigured. However, conventional pixel antennas
have only a single, fixed-position feeding port, which is not efficient
for angular sensing. Therefore, in this work, we further extend the
reconfigurability of pixel antennas by introducing the HRPA, which
enables both geometry control of the pixel antenna and switching of
its feeding ports. The model of the proposed HRPA, including both
circuit and radiation parameters, is derived. A codebook is then defined,
consisting of pixel connection states and feeding port positions for
each sensing area. Based on this codebook, an efficient optimization
approach is developed to minimize the Cram$\acute{\mathrm{\mathbf{e}}}$r-Rao
lower bound (CRLB) and obtain the optimal HRPA geometries for angular
sensing within a given area. Numerical results show that the HRPA
reduces the angle estimation error by more than 50\% across the full
three-dimensional sphere when compared with a conventional uniform
planar array of the same size. This demonstrates the effectiveness
of the proposed approach and highlights the potential of HRPA for
integrated sensing and communication systems.
\end{abstract}

\begin{IEEEkeywords}
6G, AoA, multi-port, pixel antenna, reconfigurable antenna, sensing.
\end{IEEEkeywords}

\section{Introduction}

Integrated sensing and communication (ISAC) is listed as one of the
six key usage scenarios in the sixth-generation (6G) mobile network,
which is expected to serve many novel applications such as unmanned
aerial vehicle (UAV) navigation and vehicle to everything (V2X) regulation
\cite{Zhang2019}. Angle of arrival (AoA) sensing, including azimuth
and elevation angle estimation, is one of the major tasks in ISAC
systems \cite{10530985}. Conventional multiple-input multiple-output
(MIMO) antennas, such as active antenna unit (AAU) in base stations,
employ uniform linear array (ULA) or uniform planar array (UPA) which
can be used to determine AoA according to wave path difference between
the antenna array elements \cite{Han2023}. The wave path difference
is dependent on the angular gradient of radiation patterns among antenna
elements, which is the largest at the broadside angles for ULA and
UPA. However, it degrades significantly at endfire angles. In addition,
uniform array is normally not implemented in compact devices, including
mobile phones at the user equipment (UE), due to limited size and
shape for antenna design. The irregular MIMO antennas at UE receivers
further deteriorates the angle estimation performance.

A promising technique to address these issues is utilizing the pixel
antennas which are a general reconfigurable antenna design where radiating
structure with arbitrary shape and size can be discretized into a
subwavelength elements denoted as pixels \cite{1367557}. Therefore,
pixel antennas can fully leverage the given space for antenna in compact
devices. The adjacent pixels can be connected or disconnected by using
radio frequency (RF) switches such as positive-intrinsic-negative
(PIN) diodes. By controlling the pixel connection states, the pixel
antenna topology can be flexibly adjusted so that the resulting antenna
characteristics such as radiation patterns can be reconfigured. Utilizing
this reconfigurability, various of pixel antennas have been designed
\cite{Tang2021}\nocite{Zhang2022}\nocite{Tang2023}-\nocite{Zhang2022a}\nocite{Jing2022}\nocite{10669211}\nocite{Rao2022}\nocite{Rao2023}\nocite{Zhang2021}\cite{Zhang2020}.
A function of pixel antenna is performing beam-steering where the
key advantage is that phase shifters are not needed, avoiding the
high insertion loss and reducing the cost of RF front-end when compared
to conventional uniform array \cite{Tang2021}\nocite{Zhang2022}\nocite{Tang2023}-\nocite{Zhang2022a}\nocite{Jing2022}\cite{10669211}.
Pixel antennas have also been applied to reconfigurable intelligent
surface design \cite{Rao2022}, \cite{Rao2023} and compact antennas
decoupling \cite{Zhang2021}. The cross polarization ratio of radiation
pattern can also be by using pixel antennas \cite{Zhang2020}. In
addition, pixel antennas have been utilized to enhance wireless communication.
Recently, a novel technique denoted as antenna coding has recently
been proposed \cite{Shen2024}, which refers to using a binary sequence
to control the pixel antenna configurations and resulting radiation
patterns. Consequently, the channel capacity can be improved through
enhancing channel gains or introducing additional modulation in the
beamspace \cite{Shen2024}, \cite{han2025exploiting}. Nevertheless,
these designs mainly focus on the enhancement of antenna metrics or
wireless communication capabilities, which overlooks their potential
in sensing systems. In addition, these pixel antennas have only one
feeding port which cannot efficiently perform angular sensing as conventional
multiport ULA or UPA.

Another emerging technology is the fluid antenna system (FAS) \cite{9264694},
in which the antenna can dynamically change its position within a
small range \cite{Yang2024}. The initial FAS utilizes the fluidity
of liquid metal or conductive fluid to achieve antenna movement \cite{9388928}.
However, the tuning speed and accuracy of this approach are limited.
Then the concept of FAS has been extended to reconfigurable pixel
antennas which mimic antenna movement by altering the antenna radiating
aperture \cite{10740058}. Leveraging its fluidity, the FAS can enhance
key performance metrics of MIMO communication systems, including multiplexing
gain, diversity gain, spectral efficiency, and energy efficiency \cite{New2024}-\nocite{New2024a}\nocite{Ma2024}\cite{Ye2024}.
The FAS can also assist ISAC systems by enhancing beamforming gain
towards sensing targets \cite{10705114}, \cite{10707252}. One of
the key characteristics of fluid antenna is port selection where the
RF chains can be flexibly switched to the optimal feeding ports to
adapt to the channel environment. With this property, fluid antenna
naturally fits the sensing scenarios with moving targets. Nevertheless,
an effective approach to perform angular sensing using FAS has not
yet been explored.

To address the aforementioned issues, we propose the highly reconfigurable
pixel antenna (HRPA), which integrates the advantages of pixel antennas
with extreme radiating aperture reconfigurability and FASs with highly
movable port positions. Compared with conventional angular sensing
based on uniform antenna array with fixed antenna elements, which
is sensitive to the angle of arrival (AoA) and which are rarely used
in compact devices due to size limitations, the proposed multi-port
HRPA offers flexible geometry and configurations. By switching feeding
ports and adjusting pixel connections, it can generate a set of radiation
patterns that enable accurate AoA estimation of sensing targets. The
main contributions of this paper are summarized as follows:

\textit{Firstly}, we investigate a novel antenna architecture, the
highly reconfigurable pixel antenna (HRPA), which combines both antenna
radiating aperture reconfiguration and feeding port selection for
angular sensing. The proposed multi-port HRPA can flexibly tune its
radiating aperture and feeding ports to generate radiation patterns
that adapt to the angles of sensing targets.

\textit{Secondly}, we provide the architecture and the equivalent
circuit model of HRPA. The radiation patterns of the multi-port HRPA
are derived based on multiport circuit theory as functions of the
pixel antenna geometry and port positions.

\textit{Thirdly}, we provide an efficient optimization approach to
obtain optimal geometries of HRPA by minimizing Cram$\acute{\mathrm{e}}$r-Rao
lower bound (CRLB). A corresponding codebook is then constructed for
angular sensing in different sensing areas.

\textit{Finally}, we simulate the angular sensing performance of the
HRPA. Numerical results show that by using the HRPA, angle estimation
error can be reduced by more than 50\% in the three-dimension space
compared with a conventional UPA of same size. The performance improvement
is the most significant at the endfire angles. These results demonstrate\textcolor{red}{{}
}the effectiveness of the proposed HRPA in angular sensing.

\textit{Organization}: Section II introduces the system model of angular
sensing using conventional UPA and reconfigurable pixel antennas.
Section III presents the HRPA architecture and derives its radiation
patterns using the equivalent circuit model. Section IV proposes an
optimization approach for obtaining the optimal HRPA geometries. In
Section V, the angular sensing performance of the HRPA is evaluated
to demonstrate its effectiveness. Finally, Section VI concludes the
work.

\textit{Notation}: Bold lower and upper case letters denote vectors
and matrices, respectively. Letters not in bold font represent scalars.
$\left|a\right|$, $\mathfrak{Re}\left\{ a\right\} $, $\mathfrak{Im}\left\{ a\right\} $
refers to the modulus, real part and imaginary part of a complex scalar
$a$, respectively. $\left[\mathbf{a}\right]_{i}$ and $\left\Vert \mathbf{a}\right\Vert $
refer to the $i$th entry and $l_{2}-$norm of a vector $\mathrm{\mathbf{a}}$,
respectively. $\mathrm{\mathbf{A}}^{T}$, $\mathrm{\mathbf{A}}^{H}$,
$\left[\mathrm{\mathbf{A}}\right]_{i}$ and $\left[\mathrm{\mathbf{A}}\right]_{i,j}$
refer to the transpose, conjugate transpose, $i$th column and $\left(i,j\right)$th
entry of a matrix $\mathrm{\mathbf{A}}$, respectively. $\mathbb{C}$
denotes the complex number sets and $j=\sqrt{-1}$ denotes an imaginary
number. $\mathcal{CN}(\mu,\sigma^{2})$ denotes complex Gaussian distribution
with mean $\mu$ and variance $\sigma^{2}$. $\mathbf{0}_{N}$ refers
to an $N$-dimension zero vector and $\mathbf{0}_{M\times N}$ refers
to an $M\times N$ zero matrix. $\mathbf{U}_{N}$ denotes an $N\times N$
identity matrix. diag$\left(a_{1},...,a_{N}\right)$ refers to a diagonal
matrix with diagonal elements being $a_{1},...,a_{N}$ and blkdiag$\left(\mathrm{\mathbf{A}}_{1},...,\mathrm{\mathbf{A}}_{N}\right)$
refers to a block diagonal matrix with diagonal matrices being $\mathrm{\mathbf{A}}_{1},...,\mathrm{\mathbf{A}}_{N}$.

\section{System Model \label{sec:System-Model}}

In this section, we firstly introduce the system model for angular
sensing and provide the CRLB for angle estimation. Then we describe
the angular sensing of HRPA in comparison with the conventional UPA.

\subsection{Angular Sensing Model}

Consider an angular sensing system with the sensing receiver employing
$N$ antennas, we can collect radiation patterns of the $N$ antennas
at angle of $\Omega$ as $\mathbf{E}\left(\Omega\right)=\left[\mathbf{e}_{1}\left(\Omega\right),\mathbf{e}_{2}\left(\Omega\right),...,\mathbf{e}_{N}\left(\Omega\right)\right]=\left[\mathbf{e}_{\Theta}\left(\Omega\right)\:\mathbf{e}_{\Phi}\left(\Omega\right)\right]\in\mathbb{C}^{2\times N}$
where $\mathbf{e}_{\Theta}\left(\Omega\right)=\left[e_{\Theta,1}\left(\Omega\right),e_{\Theta,2}\left(\Omega\right),...,e_{\Theta,N}\left(\Omega\right)\right]\in\mathbb{C}^{1\times N}$
and $\mathbf{e}_{\Phi}\left(\Omega\right)=\left[e_{\Phi,1}\left(\Omega\right),e_{\Phi,2}\left(\Omega\right),...,e_{\Phi,N}\left(\Omega\right)\right]\in\mathbb{C}^{1\times N}$
refer to the steering vector of the $\Theta$ and $\Phi$ polarization
components respectively. $\Omega=\left(\theta,\phi\right)$ denotes
the spatial angle with $\theta$ and $\phi$ representing the elevation
and azimuth angles in spherical coordinates, respectively. When the
angle of the sensing target with respect to the receiver, i.e., angle
of arrival (AoA), is $\Omega$, we can write the sensing system as
\begin{equation}
\mathbf{y}=\left[\begin{array}{c}
\mathbf{e}_{\Theta}\left(\Omega\right)\\
\mathbf{e}_{\Phi}\left(\Omega\right)
\end{array}\right]^{T}\left[\begin{array}{c}
s_{\Theta}\\
s_{\Phi}
\end{array}\right]+\mathbf{n},\label{eq:Signal Model}
\end{equation}
where $\mathbf{y}\in\mathbb{C}^{N\times1}$ is the received signal,
$s_{\Theta}$ and $s_{\Phi}$ are source signals for two polarizations
and can be regarded as the reflected signal from the sensing object,
$\mathbf{n}\in\mathbb{C}^{N\times1}$ is the additive Gaussian noise.
It should be noted that this model is general for both monostatic
and bistatic sensing system.

\begin{figure}[t]
\begin{centering}
\includegraphics[width=8cm]{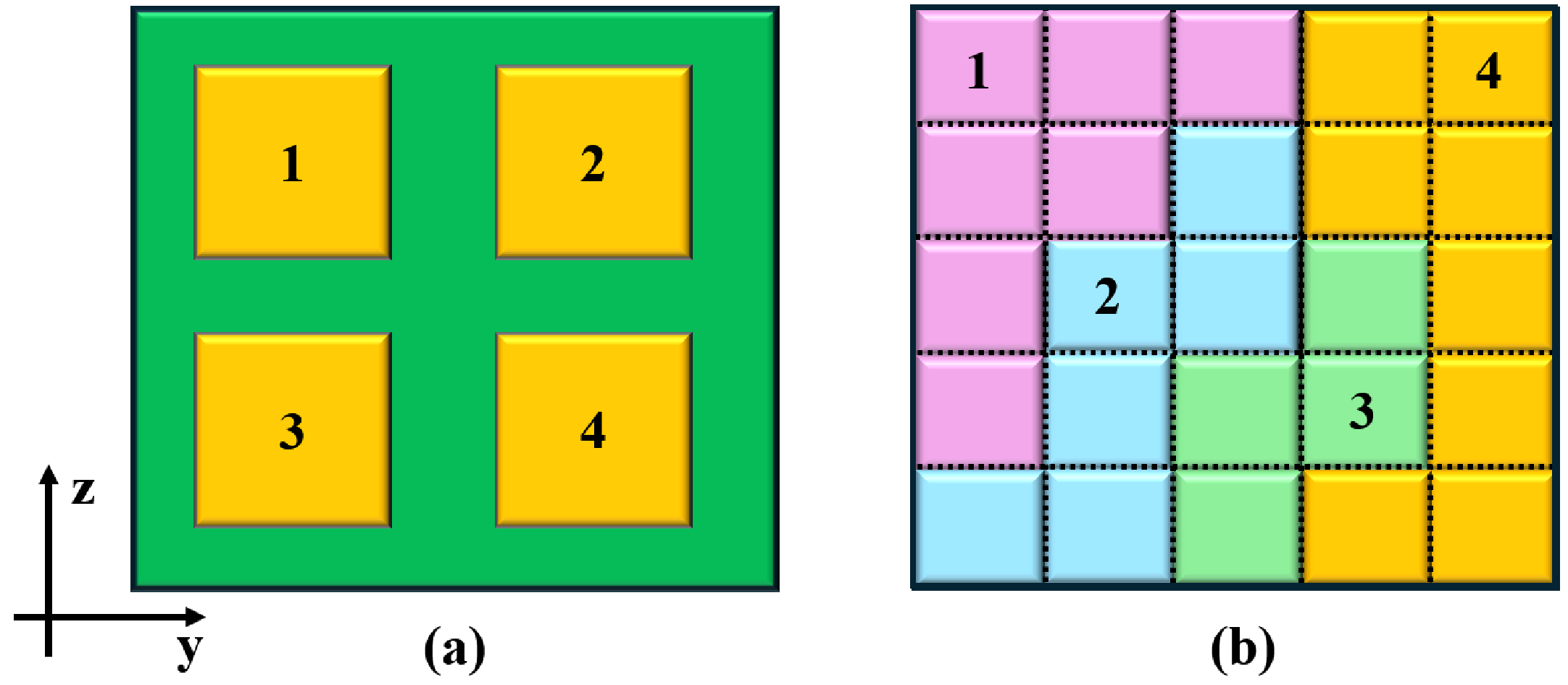}
\par\end{centering}
\caption{Illustrative examples of MIMO antennas of (a) conventional UPA and
(b) HRPA with $N=4$. \label{fig:MIMO}}
\end{figure}
The angle estimation accuracy at $\Omega$ can be evaluated by CRLB
matrix \cite{1561599}, which can be defined as
\begin{equation}
\mathbf{C}\left(\Omega\right)=\frac{\mathfrak{Re}\left\{ \mathbf{F}\left(\Omega\right)\right\} ^{-1}}{2\mathrm{SNR}}=\left[\begin{array}{cc}
c_{\theta\theta} & c_{\theta\phi}\\
c_{\phi\theta} & c_{\phi\phi}
\end{array}\right]\label{eq:CRB}
\end{equation}
where $c_{\theta\theta}$ and $c_{\phi\phi}$ refer to the CRLB for
estimation errors of $\theta$ and $\phi$ respectively, $c_{\theta\phi}$
and $c_{\phi\theta}$ refer to the CRLB for the covariance of the
estimation errors between $\theta$ and $\phi$ with $c_{\theta\phi}=c_{\phi\theta}$,
$\mathrm{SNR}$ is the signal-to-noise ratio, and $\mathbf{F}\left(\Omega\right)$
is the Fisher information matrix (FIM). Using the Slepian-Bang formula,
we write the FIM as $\mathbf{F}\left(\Omega\right)=\mathbf{J}^{H}\left(\Omega\right)\mathbf{D}\left(\Omega\right)\mathbf{J}\left(\Omega\right)$
where the Jacobian matrix $\mathbf{J}\left(\Omega\right)\in\mathbb{C}^{2K\times2}$
is given by
\begin{equation}
\mathbf{J}\left(\Omega\right)=\left[\begin{array}{cc}
\frac{\partial\mathbf{e}_{\Theta}\left(\Omega\right)}{\partial\theta} & \frac{\partial\mathbf{e}_{\Phi}\left(\Omega\right)}{\partial\theta}\\
\frac{\partial\mathbf{e}_{\Theta}\left(\Omega\right)}{\partial\phi} & \frac{\partial\mathbf{e}_{\Phi}\left(\Omega\right)}{\partial\phi}
\end{array}\right]^{T},\label{eq:Jacobian}
\end{equation}
and $\mathbf{D}$ is the projection matrix given by
\begin{equation}
\mathbf{D}\left(\Omega\right)=\mathbf{U}_{2K}-\frac{\mathbf{f}\left(\Omega\right)\mathbf{f}^{H}\left(\Omega\right)}{\left\Vert \mathbf{f}\left(\Omega\right)\right\Vert ^{2}},\label{eq:Projection}
\end{equation}
where $\mathbf{f}\left(\Omega\right)=\left[\mathbf{e}_{\Theta}\left(\Omega\right),\mathbf{e}_{\Phi}\left(\Omega\right)\right]\in\mathbb{C}^{1\times2K}$
collects steering vectors of two polarization components at angle
$\Omega$. It can be observed from \eqref{eq:CRB} to \eqref{eq:Projection}
that entries in the CRLB matrix are related to both polarization components.

\subsection{Angular Sensing}

\subsubsection{Angular Sensing by UPA}

Before introducing the HRPA, we firstly describe the angular sensing
by a conventional UPA, as illustrated in Fig. \ref{fig:MIMO} (a),
where $N$ identical antenna elements are placed in an $N_{Y}\times N_{Z}$
array with $N_{Y}N_{Z}=N$. Given the far-field radiation pattern
at spatial angle $\Omega$ of the first antenna element as $\mathbf{e}_{1}\left(\Omega\right)=\left[e_{\Theta,1}\left(\Omega\right)\:e_{\Phi,1}\left(\Omega\right)\right]{}^{T}\in\mathbb{C}^{2\times1}$,
we can write the radiation pattern of the $n$th antenna element,
$n=1,2,...,N$, using the array factor as
\begin{equation}
\text{\ensuremath{\mathbf{e}_{n}}}\left(\Omega\right)=e^{jk\left(\left(n_{y}-1\right)\mathrm{sin}\theta\mathrm{sin}\phi+\left(n_{z}-1\right)\mathrm{cos}\theta\right)}\text{\ensuremath{\mathbf{e}_{1}}}\left(\Omega\right)\label{eq:AF}
\end{equation}
where $k=\frac{2\pi d}{\lambda}$ with $d$ and $\lambda$ being the
separation distance between the adjacent antenna elements and the
wavelength, $n_{y}=n\:\mathrm{mod}\:N_{Y}$ and $n_{z}=\left\lceil \frac{n}{N_{Y}}\right\rceil $
refer to the antenna position in the UPA.

Taking \eqref{eq:AF} into \eqref{eq:CRB} to \eqref{eq:Projection},
the CRLB matrix \eqref{eq:CRB} is given by
\begin{equation}
\mathbf{C}_{\mathrm{UPA}}\left(\Omega\right)=\frac{\left[\begin{array}{cc}
b_{\mathrm{UPA},\theta\theta} & b_{\mathrm{UPA},\theta\phi}\\
b_{\mathrm{UPA},\phi\theta} & b_{\mathrm{UPA},\phi\phi}
\end{array}\right]}{2k^{4}B_{Y}B_{Z}\mathrm{sin}^{2}\theta\mathrm{cos}^{2}\phi\mathrm{SNR}}\label{eq:CRLB UPA}
\end{equation}
where $b_{\mathrm{UPA},\theta\theta}=B_{Y}\mathrm{sin}^{2}\theta\mathrm{cos}^{2}\phi$,
$b_{\mathrm{UPA},\theta\phi}=b_{\mathrm{UPA},\phi\theta}=-B_{Y}\mathrm{cos}\theta\mathrm{sin}\theta\mathrm{cos}\phi\mathrm{sin}\phi$,
and $b_{\mathrm{UPA},\phi\phi}=B_{Z}\mathrm{sin}^{2}\theta+B_{Y}\mathrm{cos}^{2}\theta\mathrm{sin}^{2}\phi$
with $B_{Y}=\frac{N_{Y}\left(N_{Y}^{2}-1\right)}{12}$ and $B_{Z}=\frac{N_{Z}\left(N_{Z}^{2}-1\right)}{12}$.
We can observe from \eqref{eq:CRLB UPA} that the angular sensing
accuracy of conventional UPA varies greatly in terms of AoA. The minimum
CRLB for estimation errors of $\theta$ and $\phi$ can be achieved
when $\theta=90^{\circ}$ and $\phi=0^{\circ}$ in the broadside direction
while CRLBs approach infinity in the endfire directions so that the
AoA of sensing targets cannot be estimated when locating on the $yoz$
plane. 

\subsubsection{Angular Sensing by HRPA}

To address the aforementioned issue, in this work, we consider using
HRPA to perform angular sensing. Compared with fixed antenna elements
and formation as in conventional UPA, HRPA is able to flexibly adjust
antenna geometries by selecting feeding ports or changing pixel connections.
As illustrated in Fig. \ref{fig:MIMO} (b), a square HRPA with the
same aperture as conventional UPA consists of $N$ different and irregular
regions (by solid lines) each of which can be viewed as one antenna
element and excited by a single feeding port. The shape and feeding
position of these $N$ elements can be changed (by dashed lines) to
various configurations. Therefore, \eqref{eq:AF} is no longer effective
for HRPA. 

We write the geometry configuration of HRPA as a variable set $\mathcal{X}$
so that we have the $N$ radiation patterns of HRPA as $\mathbf{E}\left(\mathcal{X},\Omega\right)=\left[\mathbf{e}_{1}\left(\mathcal{X},\Omega\right),...,\mathbf{e}_{N}\left(\mathcal{X},\Omega\right)\right]=\left[\mathbf{e}_{\Theta}\left(\mathcal{X},\Omega\right)\:\mathbf{e}_{\Phi}\left(\mathcal{X},\Omega\right)\right]\in\mathbb{C}^{2\times N}$
where $\text{\ensuremath{\mathbf{e}}}_{n}\left(\mathcal{X},\Omega\right)=\left[e_{\Theta,n}\left(\mathcal{X},\Omega\right)\:e_{\Phi,n}\left(\mathcal{X},\Omega\right)\right]{}^{T}\in\mathbb{C}^{2\times1}$,
$n=1,2,...,N$ denotes the radiation pattern of the $n$th feeding
port, $\mathbf{e}_{\Theta}\left(\mathcal{X},\Omega\right)=\left[e_{\Theta,1}\left(\mathcal{X},\Omega\right),e_{\Theta,2}\left(\mathcal{X},\Omega\right),...,e_{\Theta,N}\left(\mathcal{X},\Omega\right)\right]\in\mathbb{C}^{1\times N}$
and $\mathbf{e}_{\Phi}\left(\mathcal{X},\Omega\right)=\left[e_{\Phi,1}\left(\mathcal{X},\Omega\right),e_{\Phi,2}\left(\mathcal{X},\Omega\right),...,e_{\Phi,N}\left(\mathcal{X},\Omega\right)\right]\in\mathbb{C}^{1\times N}$
refer to the radiation pattern of the $\Theta$ and $\Phi$ polarization
components. The CRLB matrix for angle estimation using HRPA is thus
given by
\begin{equation}
\mathbf{C}_{\mathrm{RPA}}\left(\Omega\right)=\frac{\left[\begin{array}{cc}
b_{\mathrm{RPA},\theta\theta} & b_{\mathrm{RPA},\theta\phi}\\
b_{\mathrm{RPA},\phi\theta} & b_{\mathrm{RPA},\phi\phi}
\end{array}\right]}{2B_{\mathrm{RPA}}\mathrm{SNR}}\label{eq:CRB HRPA}
\end{equation}
where $b_{\mathrm{RPA},\theta\theta}=\left\Vert \frac{\partial\mathbf{f}_{\mathrm{RPA}}\left(\Omega\right)}{\partial\phi}\right\Vert ^{2}$
, $b_{\mathrm{RPA},\phi\phi}=\left\Vert \frac{\partial\mathbf{f}_{\mathrm{RPA}}\left(\Omega\right)}{\partial\theta}\right\Vert ^{2}$,
$b_{\mathrm{RPA},\phi\theta}=\mathfrak{Re}\left\{ \frac{\partial\mathbf{f}_{\mathrm{RPA}}\left(\Omega\right)}{\partial\theta}\frac{\partial\mathbf{f}_{\mathrm{RPA}}^{H}\left(\Omega\right)}{\partial\phi}\right\} $
with $B_{\mathrm{RPA}}=b_{\mathrm{RPA},\theta\theta}b_{\mathrm{RPA},\phi\phi}-b_{\mathrm{RPA},\theta\phi}^{2}$
and $\mathbf{f}_{\mathrm{RPA}}\left(\Omega\right)=\left[\mathbf{e}_{\Theta}\left(\mathcal{X},\Omega\right),\mathbf{e}_{\Phi}\left(\mathcal{X},\Omega\right)\right]\in\mathbb{C}^{1\times2K}$.
We can observe from \eqref{eq:CRB HRPA} that CRLBs for estimation
errors of $\theta$ and $\phi$ strongly depend on the angular gradient
of the HRPA radiation patterns. In general, the uniform antenna array
has better performance than non-uniform antenna array in terms of
the sensing area of whole 3D sphere. However for HRPA, we can optimize
the its geometry configuration for given sensing area to maximize
the angular gradient of the radiation patterns for performance enhancement.
The radiation patterns and sensing performance of HRPA will be provided
in the following sections.

\section{HRPA Model\label{sec:Fluid-Pixel-Antenna}}

In this section, we firstly describe the architecture of the HRPA.
Then we derive the radiation patterns of the HRPA, based on the equivalent
circuit model, for angular sensing. 

\subsection{Architecture of HRPA}

\begin{figure}[t]
\begin{centering}
\includegraphics[width=8.5cm]{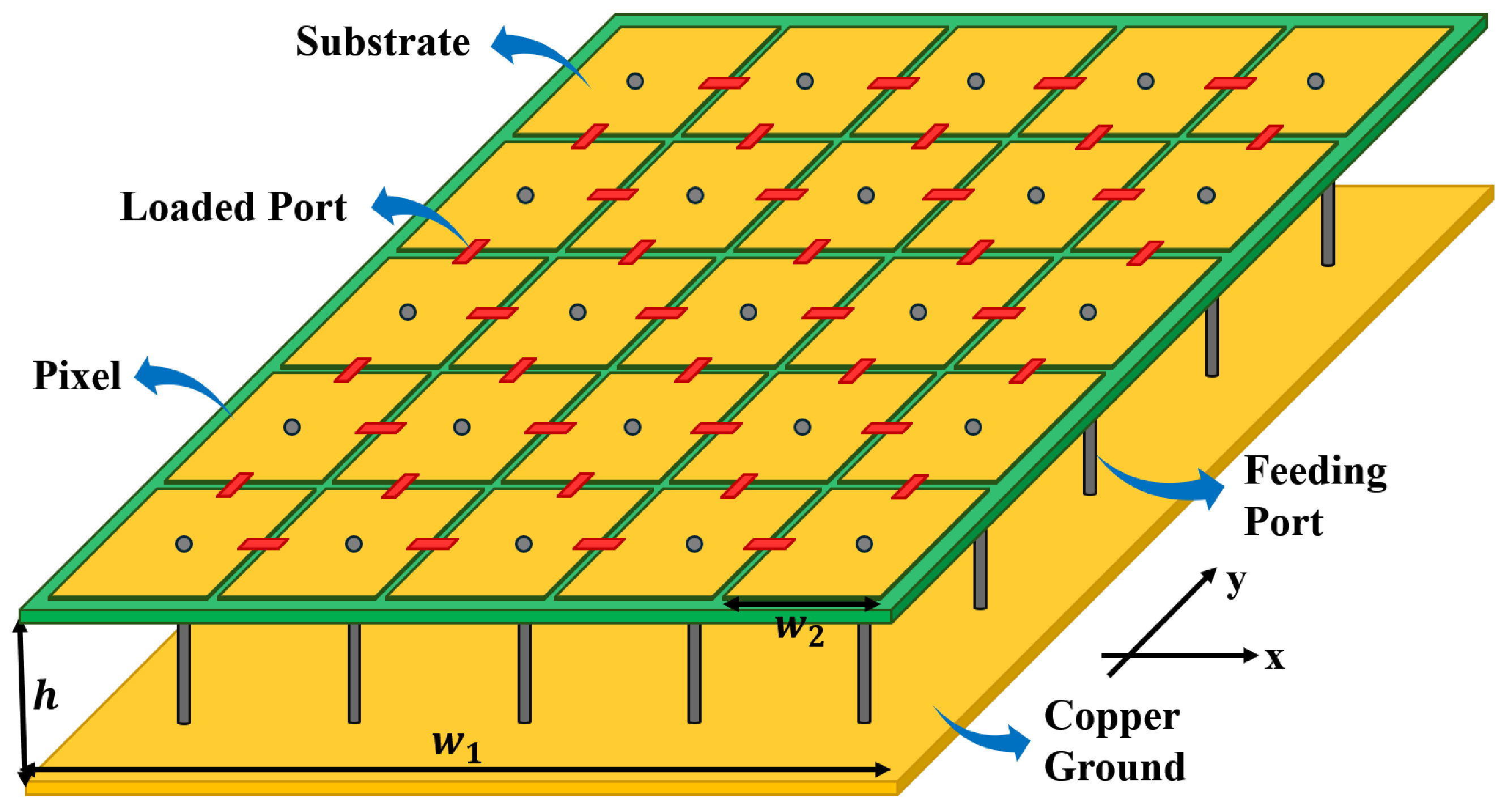}
\par\end{centering}
\begin{centering}
(a)
\par\end{centering}
\begin{centering}
\includegraphics[width=8cm]{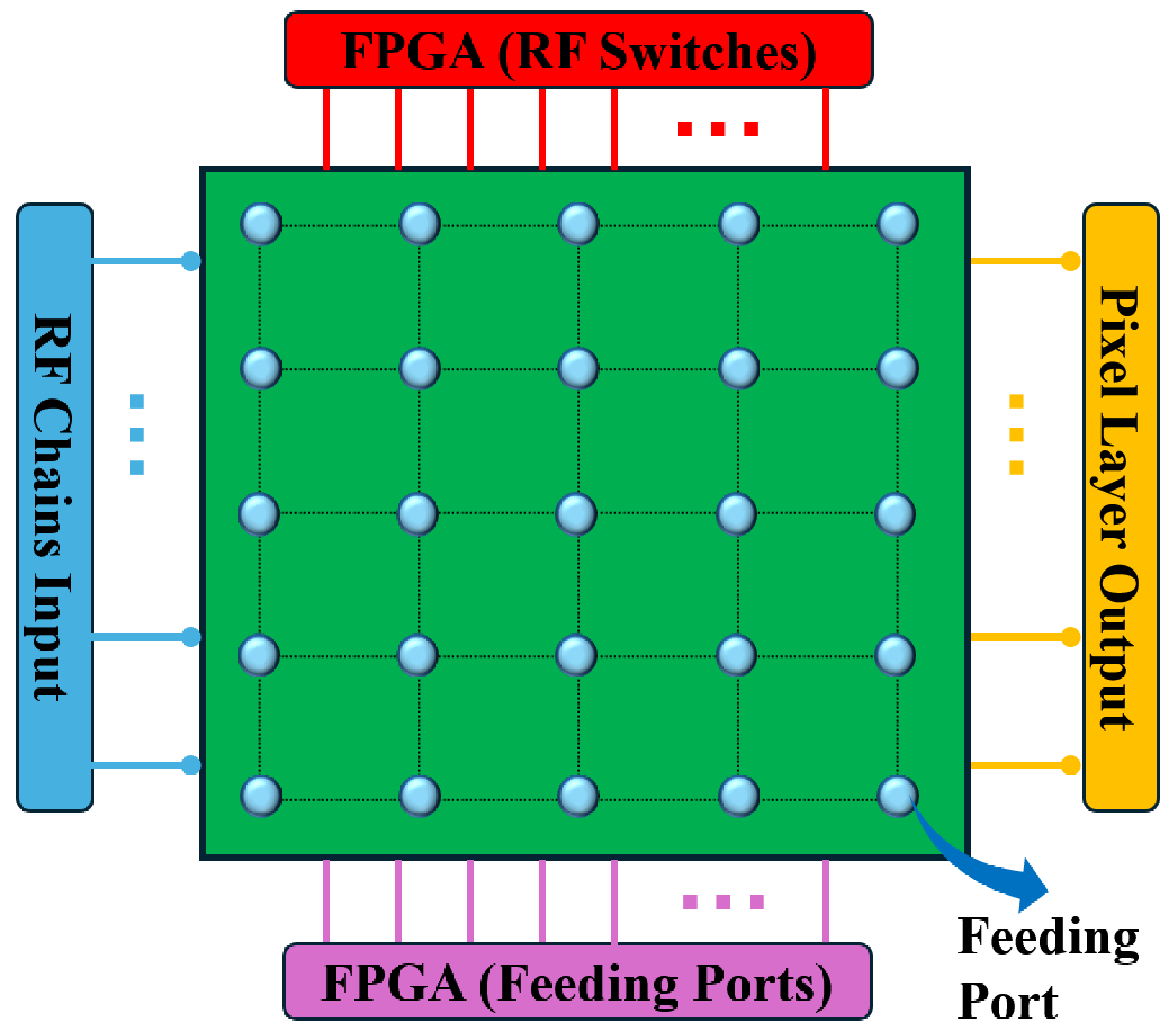}
\par\end{centering}
\begin{centering}
(b)
\par\end{centering}
\caption{(a) General architecture and (b) digitally controllable feeding port
selection module of HRPA. \label{fig:Pixel Antenna}}
\end{figure}
The architecture of the HRPA is shown in Fig. \ref{fig:Pixel Antenna}
(a), which is composed by a pixel layer, a ground layer and a feeding
port layer that connects the pixel layer and ground layer. The pixel
layer is made of a copper surface mounted on a dielectric substrate
where the copper surface is discretized into an array with sub-wavelength
elements denoted as pixels. Each pair of adjacent pixels is connected
by a loaded port which can be implemented by RF switches. By switching
on or off the RF switches, the adjacent pixels can be connected or
disconnected so that the radiating aperture of the pixel array can
be changed. Beneath the pixel layer, a copper plane is placed as the
ground so that feeding ports can be set across the ground layer and
each of the pixel centers. To control the connection states and partially
excite feeding ports, a digitally controllable module, as shown in
Fig. \ref{fig:Pixel Antenna} (b), based on the Field-Programmable
Gate Array (FPGA) is utilized. Particularly, the FPGA set up the connections
between the RF chains and the corresponding pixels, determining the
activated and muted feeding ports of the HRPA.

\begin{figure}[t]
\begin{centering}
\includegraphics[width=8.8cm]{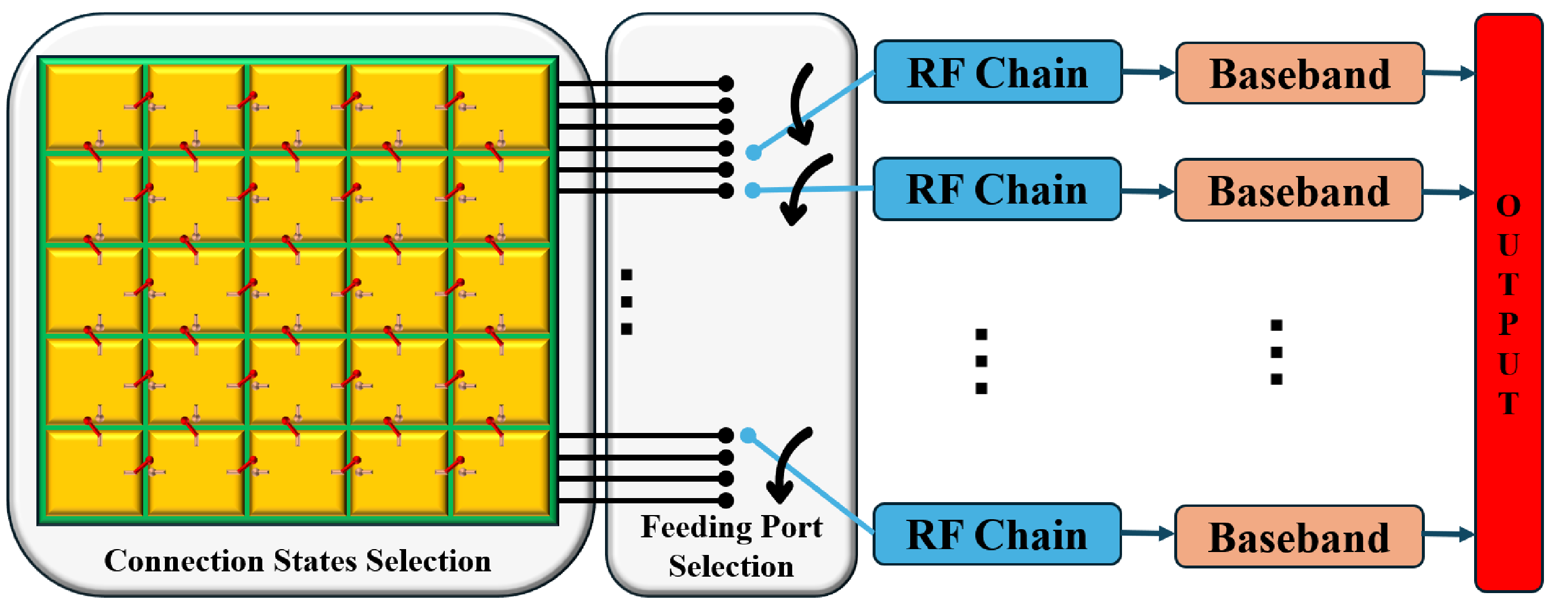}
\par\end{centering}
\begin{centering}
(b)
\par\end{centering}
\caption{MIMO sensing receiver diagram using HRPA. \label{fig:MIMO Receiver}}
\end{figure}

The overall diagram for MIMO sensing receiver based on HRPA is shown
in Fig. \ref{fig:MIMO Receiver} where the number of the activated
feeding ports is equal to the number of RF chains. The received sensing
signals are processed at the baseband for estimating the AoA. It should
be noted that this MIMO antenna diagram is a general design since
radiators of arbitrary shape and size can be discretized to form a
pixel array and the activation of feeding ports can be performed with
the digitally controllable module given arbitrary number of RF chains.

\subsection{Circuit Model of HRPA}

To systematically analyze the architecture of HRPA, we formulate the
equivalent circuit model as shown in Fig. \ref{fig:Multiport} in
which the HRPA is modeled as an $\left(M+Q\right)$-port network (indexed
by $\left\{ 1,2,...,M+Q\right\} $) characterized by a symmetric impedance
matrix $\mathbf{Z}=\mathbf{R}+j\mathbf{X}\in\mathbb{C}^{\left(M+Q\right)\times\left(M+Q\right)}$
with $\mathbf{R}=\mathfrak{Re}\left\{ \mathbf{Z}\right\} $ and $\mathbf{X}=\mathfrak{Im}\left\{ \mathbf{Z}\right\} $
being the resistance and reactance matrix respectively. In addition
in HRPA, $M$ ports are potential feeding ports (indexed by $\mathcal{M}=\left\{ 1,2,...,M\right\} $)
and $Q$ ports are loaded ports (indexed by $\mathcal{Q}=\left\{ M+1,M+2,...,M+Q\right\} $).
$N$ RF chains are connected to $N$ activated feeding ports with
the other $M-N$ ports are muted feeding ports. The open-circuit radiation
patterns of the $\left(M+Q\right)$ ports are collected into a matrix
$\mathbf{E}_{\mathrm{oc}}\left(\Omega\right)=\left[\mathrm{\mathbf{e}}_{\mathrm{oc},1}\left(\Omega\right),\mathrm{\mathbf{e}}_{\mathrm{oc},2}\left(\Omega\right),...,\mathrm{\mathbf{e}}_{\mathrm{oc},M+Q}\left(\Omega\right)\right]=\left[\mathbf{e}_{\mathrm{oc},\Theta}\left(\Omega\right)\:\mathbf{e}_{\mathrm{oc},\Phi}\left(\Omega\right)\right]\in\mathbb{C}^{2\times\left(M+Q\right)}$
with $\mathbf{e}_{\mathrm{oc},\Theta}\left(\Omega\right)\in\mathbb{C}^{1\times\left(M+Q\right)}$
and $\mathbf{e}_{\mathrm{oc},\Phi}\left(\Omega\right)\in\mathbb{C}^{1\times\left(M+Q\right)}$
denoting the open-circuit radiation patterns of the $\Theta$ and
$\Phi$ polarization components respectively. The analysis of the
equivalent circuit model is provided in this subsection.

\subsubsection{Port Selection}

\begin{figure}[t]
\begin{centering}
\includegraphics[width=8.8cm]{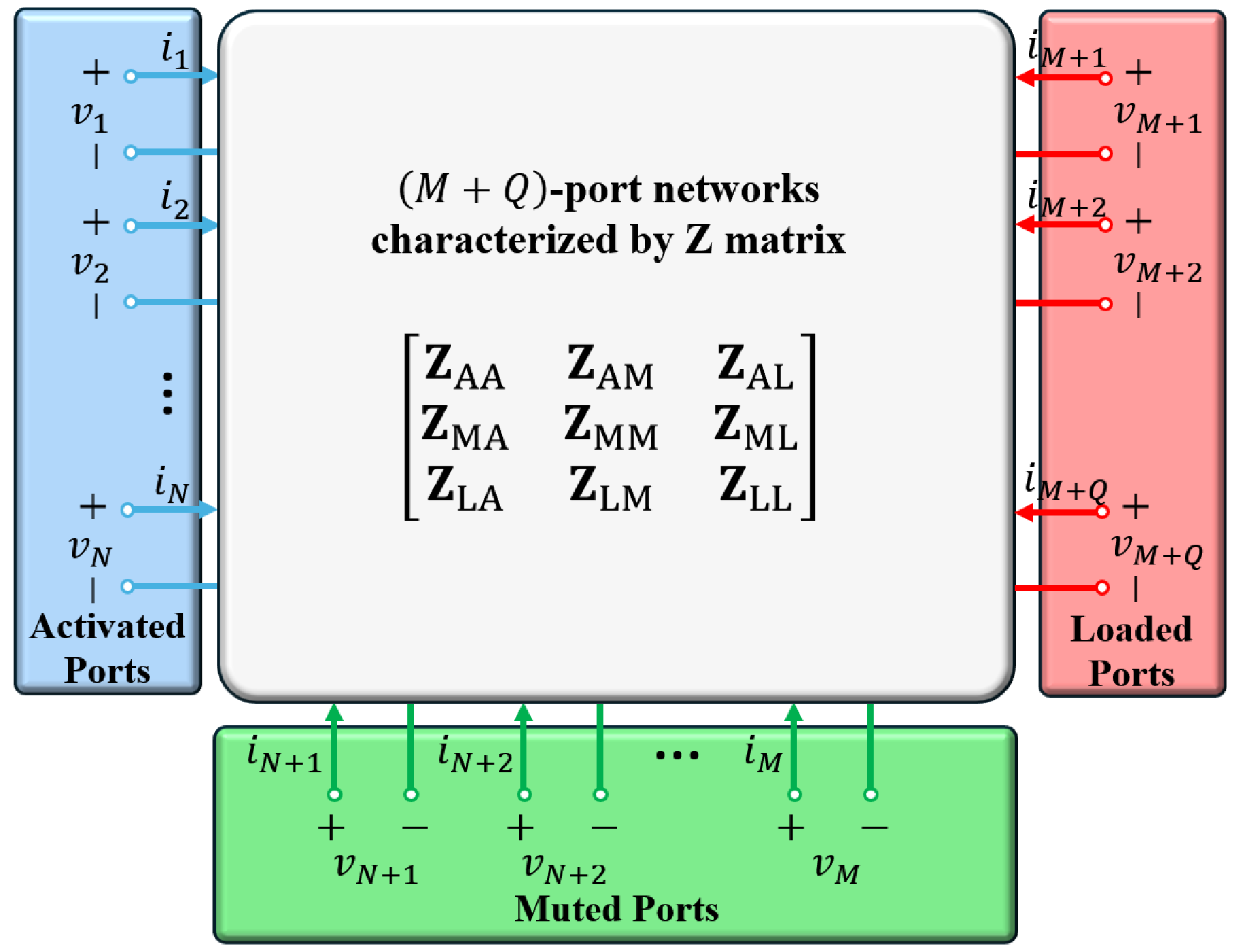}
\par\end{centering}
\caption{Equivalent circuit model of $\left(M+Q\right)$-port network for HRPA
with $N$ activated feeding ports, $M-N$ muted feeding ports and
$Q$ loaded ports. \label{fig:Multiport}}
\end{figure}
We start from modeling the feeding port selection of HRPA. Defining
the position indices set, $\mathcal{F}=\left\{ f_{1},f_{2},...,f_{N}\right\} $,
which consists of $N$ different indices of activated feeding ports
$f_{n}$ with $1\leqslant f_{n}\leqslant M$, $n=1,2,...,N$, we can
write a permutation matrix for feeding port selection given by
\begin{equation}
\mathbf{P}\left(\mathcal{F}\right)=\left[\mathbf{P}_{\mathrm{A}}\left(\mathcal{F}\right),\mathbf{P}_{\mathrm{M}}\left(\mathcal{F}\right),\mathbf{P}_{\mathrm{L}}\right]\in\mathbb{C}^{\left(M+Q\right)\times\left(M+Q\right)}
\end{equation}
where the $f_{n}$th entry in the $n$th column of $\mathbf{P}_{\mathrm{A}}\left(\mathcal{F}\right)\in\mathbb{C}^{\left(M+Q\right)\times N}$
is unity with the other entries being zero, which is used to select
the $f_{n}$th port as the activated feeding port. In addition, $\mathbf{P}_{\mathrm{L}}=\left[\mathbf{0}_{M\times Q},\mathbf{U}_{Q}\right]^{T}\in\mathbb{C}^{\left(M+Q\right)\times Q}$
where the $q$th entry, $q=M+1,M+2,..,M+Q$, in the $\left(q-M\right)$th
column of is unity with the other entries being zero, which denotes
the loaded port index and is irrelevant to the feeding port indices
as shown in Fig. \ref{fig:Multiport}. While each column in $\mathbf{P}_{\mathrm{M}}\left(\mathcal{F}\right)$
contains a single unity entry whose position index is included in
$\mathcal{M}\setminus\mathcal{F}$, indicating the port indices of
the muted feeding ports. Using the permutation matrix $\mathbf{P}\left(\mathcal{F}\right)$
to select activated feeding ports, we can rearrange the impedance
matrix $\mathbf{Z}$ as
\begin{equation}
\mathbf{P}^{T}\left(\mathcal{F}\right)\mathbf{Z}\mathbf{P}\left(\mathcal{F}\right)=\left[\begin{array}{ccc}
\mathbf{Z}_{\mathrm{AA}}\left(\mathcal{F}\right) & \mathbf{Z}_{\mathrm{AM}}\left(\mathcal{F}\right) & \mathbf{Z}_{\mathrm{AL}}\left(\mathcal{F}\right)\\
\mathbf{Z}_{\mathrm{MA}}\left(\mathcal{F}\right) & \mathbf{Z}_{\mathrm{MM}}\left(\mathcal{F}\right) & \mathbf{Z}_{\mathrm{ML}}\left(\mathcal{F}\right)\\
\mathbf{Z}_{\mathrm{LA}}\left(\mathcal{F}\right) & \mathbf{Z}_{\mathrm{LM}}\left(\mathcal{F}\right) & \mathbf{Z}_{\mathrm{LL}}
\end{array}\right],\label{eq:Impedance}
\end{equation}
where $\mathbf{Z}_{\mathrm{AA}}\left(\mathcal{F}\right)=\mathbf{P}_{\mathrm{A}}^{T}\left(\mathcal{F}\right)\mathbf{Z}\mathbf{P}_{\mathrm{A}}\left(\mathcal{F}\right)\in\mathbb{C}^{N\times N}$,
$\mathbf{Z}_{\mathrm{MM}}\left(\mathcal{F}\right)=\mathbf{P}_{\mathrm{M}}^{T}\left(\mathcal{F}\right)\mathbf{Z}\mathbf{P}_{\mathrm{M}}\left(\mathcal{F}\right)\in\mathbb{C}^{\left(M-N\right)\times\left(M-N\right)}$
and $\mathbf{Z}_{\mathrm{LL}}=\mathbf{P}_{\mathrm{L}}^{T}\mathbf{Z}\mathbf{P}_{\mathrm{L}}\in\mathbb{C}^{Q\times Q}$
are the self impedance sub-matrices of the $N$ activated feeding
ports, $M-N$ muted feeding ports and $Q$ loaded ports, respectively.
$\mathbf{Z}_{\mathrm{AM}}\left(\mathcal{F}\right)=\mathbf{P}_{\mathrm{A}}^{T}\left(\mathcal{F}\right)\mathbf{Z}\mathbf{P}_{\mathrm{M}}\left(\mathcal{F}\right)\in\mathbb{C}^{N\times\left(M-N\right)}$,
$\mathbf{Z}_{\mathrm{AL}}\left(\mathcal{F}\right)=\mathbf{P}_{\mathrm{A}}^{T}\left(\mathcal{F}\right)\mathbf{Z}\mathbf{P}_{\mathrm{L}}\in\mathbb{C}^{N\times Q}$
and $\mathbf{Z}_{\mathrm{ML}}\left(\mathcal{F}\right)=\mathbf{P}_{\mathrm{M}}^{T}\left(\mathcal{F}\right)\mathbf{Z}\mathbf{P}_{\mathrm{L}}\in\mathbb{C}^{\left(M-N\right)\times Q}$
are the mutual impedance matrices between all pairs among the activated
feeding ports, muted feeding ports and loaded ports respectively with
$\mathbf{Z}_{\mathrm{AM}}=\mathbf{Z}_{\mathrm{MA}}^{T}$, $\mathbf{Z}_{\mathrm{AL}}=\mathbf{Z}_{\mathrm{LA}}^{T}$
and $\mathbf{Z}_{\mathrm{ML}}=\mathbf{Z}_{\mathrm{LM}}^{T}$. 

Correspondingly, we can also use the permutation matrix to obtain
the open-circuit radiation patterns of the activated feeding ports,
muted feeding ports and loaded ports as $\mathbf{E}_{\mathrm{A}}\left(\mathcal{F},\Omega\right)=\mathbf{E}_{\mathrm{oc}}\left(\Omega\right)\mathbf{P}_{\mathrm{A}}\left(\mathcal{F}\right)$,
$\mathbf{E}_{\mathrm{M}}\left(\mathcal{F},\Omega\right)=\mathbf{E}_{\mathrm{oc}}\left(\Omega\right)\mathbf{P}_{\mathrm{M}}\left(\mathcal{F}\right)$
and $\mathbf{E}_{\mathrm{L}}\left(\Omega\right)=\mathbf{E}_{\mathrm{oc}}\left(\Omega\right)\mathbf{P}_{\mathrm{L}}$. 

\subsubsection{Pixel Connections}

Then we model the pixel connection states for the HRPA. A loaded port
is placed across each pair of adjacent pixels, which can be implemented
by an RF switch. Each RF switch has two states, i.e., switch on and
off states, which can be denoted as a binary variable $g_{q}\in\left\{ 0,1\right\} $
for $q=M+1,M+2,..,M+Q$. The corresponding load impedance can be either
open- or short-circuit, written as \cite{Shen2024}
\begin{equation}
z_{\mathrm{L},q}=\begin{cases}
\infty, & g_{q}=1,\:\textrm{i.e. switch\:off},\\
0, & g_{q}=0,\:\textrm{i.e. switch\:on},
\end{cases}\label{eq:Switch-1}
\end{equation}
where numerically we can use a very large value $z_{\mathrm{oc}}$
to approximate the open-circuit load impedance $\infty$. Namely,
very large load impedance can make adjacent pixels open-circuited
that is equivalent to switching off, while zero load impedance can
make adjacent pixel short-circuit that is equivalent to switching
on. We collect $g_{q}$ into a binary vector $\mathbf{g}=\left[g_{M+1},g_{M+2},....,g_{M+Q}\right]^{T}\in\mathbb{\mathbb{R}}^{Q\times1}$
which controls the pixel connection states of HRPA. Accordingly, we
can write the load impedance for all $Q$ loaded ports as a diagonal
matrix
\begin{align}
\mathbf{Z}_{\mathrm{L}}^{L}\left(\mathbf{g}\right) & =\mathrm{diag}\left(z_{\mathrm{L},M+1},z_{\mathrm{L},M+2},...,z_{\mathrm{L},M+Q}\right)\nonumber \\
 & =z_{\mathrm{oc}}\mathrm{diag}\left(g_{M+1},g_{M+2},....,g_{M+Q}\right).\label{eq:Load Matrix}
\end{align}

\subsubsection{Overall HRPA Model}

With the open-circuit radiation patterns and impedance matrices, we
aim to derive the closed-form expressions for the multi-port HRPA
using the multi-port circuit theory. The voltage source excitation
at the $N$ active feeding ports $v_{n}$, $n=1,2,...,N$, can be
grouped into a vector as $\mathbf{v}_{\mathrm{A}}=\left[v_{1},v_{2},...,v_{N}\right]{}^{T}\in\mathbb{C}^{N\times1}$
while the muted feeding ports are open-circuit which are equivalent
to be implemented by zero source voltage $\mathbf{v}_{\mathrm{M}}=\mathbf{0}_{M-N}$
and infinite inner impedance $\mathbf{Z}_{\mathrm{M}}^{L}=\mathrm{diag}\left(\infty,\infty,...,\infty\right)\in\mathbb{C}^{\left(M-N\right)\times\left(M-N\right)}$.
The $Q$ loaded ports are equipped with zero source voltage $\mathbf{v}_{\mathrm{L}}=\mathbf{0}_{Q}$.
In addition, we also group currents at the activated feeding ports,
muted feeding ports and loaded ports into vectors as $\mathbf{i}_{\mathrm{A}}=\left[i_{1},i_{2},...,i_{N}\right]{}^{T}\in\mathbb{C}^{N\times1}$,
$\mathbf{i}_{\mathrm{M}}=\left[i_{N+1},i_{N+2},...,i_{M}\right]{}^{T}\in\mathbb{C}^{\left(M-N\right)\times1}$
and $\mathbf{i}_{\mathrm{L}}=\left[i_{M+1},i_{M+2},...,i_{M+Q}\right]{}^{T}\in\mathbb{C}^{Q\times1}$
respectively. Thus we can relate the voltage and current in the $\left(M+Q\right)$-port
network as shown in Fig. \ref{fig:Multiport} (the dependence on $\mathcal{F}$
and $\mathbf{g}$ are not shown for brevity) by
\begin{align}
\left[\begin{array}{c}
\mathbf{v}_{\mathrm{A}}\\
\mathbf{v}_{\mathrm{M}}\\
\mathbf{v}_{\mathrm{L}}
\end{array}\right] & =\left(\mathbf{P}^{T}\mathbf{Z}\mathbf{P}+\mathrm{blkdiag}\left(\mathbf{0},\mathbf{Z}_{\mathrm{M}}^{L},\mathbf{Z}_{\mathrm{L}}^{L}\right)\right)\left[\begin{array}{c}
\mathbf{i}_{\mathrm{A}}\\
\mathbf{i}_{\mathrm{M}}\\
\mathbf{i}_{\mathrm{L}}
\end{array}\right]\nonumber \\
 & =\left[\begin{array}{ccc}
\mathbf{Z}_{\mathrm{AA}} & \mathbf{Z}_{\mathrm{AM}} & \mathbf{Z}_{\mathrm{AL}}\\
\mathbf{Z}_{\mathrm{MA}} & \mathbf{Z}_{\mathrm{MM}}+\mathbf{Z}_{\mathrm{M}}^{L} & \mathbf{Z}_{\mathrm{ML}}\\
\mathbf{Z}_{\mathrm{LA}} & \mathbf{Z}_{\mathrm{LM}} & \mathbf{Z}_{\mathrm{LL}}+\mathbf{Z}_{\mathrm{L}}^{L}
\end{array}\right]\left[\begin{array}{c}
\mathbf{i}_{\mathrm{A}}\\
\mathbf{i}_{\mathrm{M}}\\
\mathbf{i}_{\mathrm{L}}
\end{array}\right].\label{eq:V=00003DZI}
\end{align}
From \eqref{eq:V=00003DZI} we can obtain the relationship between
the current at the activated feeding ports and the other ports as
\begin{equation}
\left[\begin{array}{c}
\mathbf{i}_{\mathrm{M}}\\
\mathbf{i}_{\mathrm{L}}
\end{array}\right]=-\left[\begin{array}{cc}
\mathbf{Z}_{\mathrm{MM}}+\mathbf{Z}_{\mathrm{M}}^{L} & \mathbf{Z}_{\mathrm{ML}}\\
\mathbf{Z}_{\mathrm{LM}} & \mathbf{Z}_{\mathrm{LL}}+\mathbf{Z}_{\mathrm{L}}^{L}
\end{array}\right]^{-1}\left[\begin{array}{c}
\mathbf{Z}_{\mathrm{MA}}\\
\mathbf{Z}_{\mathrm{LA}}
\end{array}\right]\mathbf{i}_{\mathrm{A}}.\label{eq:current relation}
\end{equation}
Since $\mathbf{Z}_{\mathrm{M}}^{L}$ is a diagonal matrix with diagonal
entries being infinity, \eqref{eq:current relation} can be approximately
derived as
\begin{align}
\left[\begin{array}{c}
\mathbf{i}_{\mathrm{M}}\\
\mathbf{i}_{\mathrm{L}}
\end{array}\right] & =\left[\begin{array}{cc}
\mathbf{0}_{\left(M-N\right)\times\left(M-N\right)} & \mathbf{0}_{\left(M-N\right)\times Q}\\
\mathbf{0}_{Q\times\left(M-N\right)} & -\left(\mathbf{Z}_{\mathrm{LL}}+\mathbf{Z}_{\mathrm{L}}^{L}\right)^{-1}
\end{array}\right]\left[\begin{array}{c}
\mathbf{Z}_{\mathrm{MA}}\\
\mathbf{Z}_{\mathrm{LA}}
\end{array}\right]\mathbf{i}_{\mathrm{A}}\nonumber \\
 & =\left[\begin{array}{c}
\mathbf{0}_{\left(M-N\right)\times1}\\
\left(\mathbf{Z}_{\mathrm{LL}}+\mathbf{Z}_{\mathrm{L}}^{L}\right)^{-1}\mathbf{Z}_{\mathrm{LA}}\mathbf{i}_{\mathrm{A}}
\end{array}\right].\label{eq:current approx}
\end{align}
Using \eqref{eq:current approx}, the impedance matrix among $N$
activated feeding ports (selected by $\mathcal{F}$) when with pixel
connection states $\mathbf{g}$ can be written as
\begin{align}
\mathbf{Z}_{\mathrm{F}}\left(\mathcal{F},\mathbf{g}\right) & =\mathbf{R}_{\mathrm{F}}\left(\mathcal{F},\mathbf{g}\right)+j\mathbf{X}_{\mathrm{F}}\left(\mathcal{F},\mathbf{g}\right)\nonumber \\
 & =\mathbf{Z}_{\mathrm{AA}}-\left[\mathbf{Z}_{\mathrm{AM}},\mathbf{Z}_{\mathrm{AL}}\right]\cdot\left[\begin{array}{c}
\mathbf{i}_{\mathrm{M}}\\
\mathbf{i}_{\mathrm{L}}
\end{array}\right]\nonumber \\
 & =\mathbf{Z}_{\mathrm{AA}}-\mathbf{Z}_{\mathrm{AL}}\left(\mathbf{Z}_{\mathrm{LL}}+\mathbf{Z}_{\mathrm{L}}^{L}\right)^{-1}\mathbf{Z}_{\mathrm{LA}}\label{eq:Z feed}
\end{align}
where $\mathbf{R}_{\mathrm{F}}\left(\mathcal{F},\mathbf{g}\right)=\mathfrak{Re}\left\{ \mathbf{Z}_{\mathrm{F}}\left(\mathcal{F},\mathbf{g}\right)\right\} $
and $\mathbf{X}_{\mathrm{F}}\left(\mathcal{F},\mathbf{g}\right)=\mathfrak{Im}\left\{ \mathbf{Z}_{\mathrm{F}}\left(\mathcal{F},\mathbf{g}\right)\right\} $
denote the resistance and reactance matrix among $N$ activated feeding
ports. It should be noted that $\mathbf{Z}_{\mathrm{AA}}$, $\mathbf{Z}_{\mathrm{AL}}$
and $\mathbf{Z}_{\mathrm{LA}}$ are associated to the activated feeding
port indices $\mathcal{F}$ while $\mathbf{Z}_{\mathrm{LL}}$ is determined
by the pixel connection states $\mathbf{g}$. Correspondingly, the
open-circuit radiation patterns of $N$ activated feeding ports with
pixel connection states $\mathbf{g}$ can be obtained as
\begin{align}
\mathbf{E}_{\mathrm{oc,F}}\left(\mathcal{F},\mathbf{g},\Omega\right)= & \mathbf{E}_{\mathrm{A}}-\left[\mathbf{E}_{\mathrm{M}},\mathbf{E}_{\mathrm{L}}\right]\cdot\nonumber \\
 & \left[\begin{array}{cc}
\mathbf{Z}_{\mathrm{MM}}+\mathbf{Z}_{\mathrm{M}}^{L} & \mathbf{Z}_{\mathrm{ML}}\\
\mathbf{Z}_{\mathrm{LM}} & \mathbf{Z}_{\mathrm{LL}}+\mathbf{Z}_{\mathrm{L}}^{L}
\end{array}\right]^{-1}\left[\begin{array}{c}
\mathbf{Z}_{\mathrm{MA}}\\
\mathbf{Z}_{\mathrm{LA}}
\end{array}\right]\nonumber \\
= & \mathbf{E}_{\mathrm{oc}}\left(\mathbf{P}_{\mathrm{A}}-\mathbf{P}_{\mathrm{L}}\left(\mathbf{Z}_{\mathrm{LL}}+\mathbf{Z}_{\mathrm{L}}^{L}\right)^{-1}\mathbf{Z}_{\mathrm{LA}}\right).\label{eq:F pattern}
\end{align}
Given the inner impedances of sources at the $N$ active feeding ports
$\mathbf{Z}_{0}=\mathrm{diag}\left(z_{0,1},z_{0,2},...,z_{0,N}\right)$
with $z_{0,n}$, $n=1,2,...,N$, denoting the inner impedance of source
connected to the $n$th activated feeding ports, the radiation patterns
of the $N$ activated feeding ports, including the mutual coupling
effect, can be obtained as \cite{Zhang2021}
\begin{equation}
\mathbf{E}_{\mathrm{F}}\left(\mathcal{F},\mathbf{g},\Omega\right)=\mathbf{E}_{\mathrm{oc,F}}\left(\mathcal{F},\mathbf{g},\Omega\right)\left(\mathbf{Z}_{0}+\mathbf{Z}_{\mathrm{F}}\left(\mathcal{F},\mathbf{g}\right)\right)^{-1},\label{eq:Pattern Coupling}
\end{equation}
where $\mathbf{E}_{\mathrm{F}}\left(\mathcal{F},\mathbf{g},\Omega\right)=\left[\mathbf{e}_{\mathrm{F},\Theta}\left(\mathcal{F},\mathbf{g},\Omega\right)\:\mathbf{e}_{\mathrm{F},\Phi}\left(\mathcal{F},\mathbf{g},\Omega\right)\right]=\left[\mathrm{\mathbf{e}}_{\mathrm{F},1}\left(\mathcal{F},\mathbf{g},\Omega\right),...,\mathrm{\mathbf{e}}_{\mathrm{F},N}\left(\mathcal{F},\mathbf{g},\Omega\right)\right]\in\mathbb{C}^{2\times N}$.
In addition, the radiation efficiencies of the $N$ activated feeding
ports can be written as a diagonal matrix $\Lambda\left(\mathcal{F},\mathbf{g}\right)=\mathrm{diag}\left(\lambda_{1},\lambda_{2},...,\lambda_{N}\right)$
with diagonal entry $\lambda_{n}$ referring to the radiation efficiency
of the $n$th activated feeding port given by
\begin{align}
\lambda_{n}\left(\mathcal{F},\mathbf{g}\right) & =\frac{\int_{\phi=-\pi}^{\pi}\int_{\theta=0}^{\pi}\mathrm{\mathbf{e}}_{\mathrm{F},n}^{H}\mathrm{\mathbf{e}}_{\mathrm{F},n}\mathrm{d}\theta\mathrm{d}\phi}{\eta\mathfrak{Re}\left\{ \left[\mathbf{v}_{\mathrm{A}}^{H}\right]_{n}\left[\left(\mathbf{Z}_{0}+\mathbf{Z}_{\mathrm{F}}\right)^{-1}\mathbf{v}_{\mathrm{A}}\right]_{n}\right\} }\label{eq:Rad Eff}
\end{align}
where $\eta=120\pi$ is the intrinsic impedance of free space. With
\eqref{eq:Pattern Coupling} and \eqref{eq:Rad Eff}, the overall
radiation patterns of the proposed HRPA for angular sensing can be
obtained as
\begin{equation}
\mathbf{E}\left(\mathcal{F},\mathbf{g},\Omega\right)=\mathbf{E}_{\mathrm{F}}\left(\mathcal{F},\mathbf{g},\Omega\right)\Lambda^{\frac{1}{2}}\left(\mathcal{F},\mathbf{g}\right)\label{eq:Overall Pattern}
\end{equation}
which includes the effect of both mutual impedances and radiation
patterns of $N$ active feeding ports of HRPA. As a result, by taking
\eqref{eq:Pattern Coupling} into \eqref{eq:CRB HRPA} and defining
variable set as $\mathcal{X}=\left(\mathcal{F},\mathbf{g}\right)$,
we can obtain CRLB of HRPA, denoted as $\mathbf{C}_{\mathrm{RPA}}\left(\mathcal{F},\mathbf{g},\Omega\right)$,
to evaluate the angle estimation performance at a given spatial angle
$\Omega$.

We can use a full electromagnetic solver, CST studio suite \cite{CMS},
to simulate the impedance matrix $\mathbf{Z}$ and open-circuit radiation
patterns $\mathbf{E}_{\mathrm{oc}}\left(\Omega\right)$ of all $\left(M+Q\right)$
ports of HRPA. This only needs to be performed once since any impedance
matrix and radiation patterns of activated feeding ports can then
be found by using \eqref{eq:Z feed} and \eqref{eq:F pattern} \cite{Zhang2021}.

\section{Optimization}

To obtain the optimal geometries of the proposed HRPA for angular
sensing, we wish to select the $N$ active feeding ports among $M$
potential feeding ports and optimize the pixel connection states of
$Q$ loaded ports so that CRLB of angular sensing errors in given
areas can be minimized. Leveraging the CRLB matrix \eqref{eq:CRB},
we can formulate the optimization problem to find the HRPA geometry
configuration as
\begin{align}
\underset{\mathcal{F},\mathbf{g}}{\mathrm{min}}\:\:\mathrm{max}\:\: & \sqrt{\mathrm{Tr}\left(\mathbf{C}_{\mathrm{RPA}}\left(\mathcal{F},\mathbf{g},\Omega\right)\right)},\label{eq:problem}\\
\mathrm{s.t.}\:\: & \mathrm{card}\left(\mathcal{F}\right)=N,\label{eq:Cons1}\\
 & \mathbf{g}\in\left\{ 0,1\right\} ^{Q}.\label{eq:Cons2}
\end{align}
The objective function can evaluate the overall performance of solid
angle estimation and aims to simultaneously minimize the maximum CRLB
of azimuth and elevation angle estimation for any angle in the sensing
area $\Omega$. Problem \eqref{eq:problem} to \eqref{eq:Cons2} involves
two types of variables, i.e., activated feeding port indices $\mathcal{F}$
and pixel connection states $\mathbf{g}$. There is totally $\binom{M}{N}\times2^{Q}$
possible geometries of HRPA, including $2^{Q}$ combinations of pixel
connection states and $\binom{M}{N}$ combinations of activated feeding
port indices. Hence, it is impossible to search through all geometries.
Besides, the expression of the objective function is also complicated
and non-convex. Specifically, altering the activated feeding port
indices significantly changes the impedance sub-matrices in \eqref{eq:Impedance}
and open-circuit radiation patterns in \eqref{eq:F pattern} while
finding the optimal pixel connection states is a binary optimization
problem. For these reasons, it is complicated to solve problem \eqref{eq:problem}
to \eqref{eq:Cons2}.

To address this issue, we propose an efficient optimization approach,
based on alternating optimization method, to iteratively optimize
pixel connection states and select activated feeding ports. To start
with, the initial guess of the activated feeding port indices and
pixel connection states at iteration 0 are denoted as $\mathcal{F}^{\left(0\right)}$
and $\mathbf{g}^{\left(0\right)}$. At the $i$th iteration, we firstly
optimize the pixel connection states $\mathbf{g}^{\left(i\right)}$,
with fixed activated feeding port indices $\mathcal{F}^{\left(i-1\right)}$
and the resulting permutation matrix $\mathbf{P}\left(\mathcal{F}^{\left(i-1\right)}\right)$,
by solving the optimization problem
\begin{align}
\underset{\mathbf{g}^{\left(i\right)}}{\mathrm{min}}\:\:\mathrm{max}\:\: & \sqrt{\mathrm{Tr}\left(\mathbf{C}_{\mathrm{RPA}}\left(\mathcal{F}^{\left(i-1\right)},\mathbf{g}^{\left(i\right)},\Omega\right)\right)},\label{eq:problem-1}\\
\mathrm{s.t.}\:\: & \mathbf{g}^{\left(i\right)}\in\left\{ 0,1\right\} ^{Q}.\label{eq:Cons1-1}
\end{align}
which is a binary optimization problem. Multiple optimization methods
including successive exhaustive Boolean optimization \cite{7762757},
perturbation sensitivity \cite{9491941} and adjoint method \cite{10035928}
have been proposed. In this work, we use the genetic algorithm (GA)
to solve the problem \eqref{eq:problem-1}-\eqref{eq:Cons1-1}. The
optimal pixel connection states are obtained as $\mathbf{g}^{\left(i\right)}$\cite{Zhang2022}.

\begin{algorithm}[t]
\caption{The alternating optimization method.\label{alg:AO}}

\textbf{Input:} $\mathcal{X}_{\left(0\right),1}^{\star}=\left(\mathcal{F}_{\left(0\right),1}^{\star},\mathbf{g}_{\left(0\right),1}^{\star}\right)$,
$\Omega_{\left(0\right)}$, $K_{t}$, $t=1,...,T$;

$\quad${\scriptsize 1:} \textbf{for} $t=1\colon T$

$\quad${\scriptsize 2:} $\:\:$Equally divide $\Omega_{\left(t-1\right),k}$
into $K_{t}$ areas

$\quad$$\quad$$\:\:$$\Omega_{\left(t\right),k}$, $k=1,...,\prod_{t'=1}^{t}K_{t'}$;

$\quad${\scriptsize 3:} $\:\:$\textbf{for} $k=1\colon\prod_{t'=1}^{t}K_{t'}$

$\quad${\scriptsize 4:} $\:\:$$\:\:$\textbf{Initialization:} $\mathcal{F}_{\left(t\right),k}^{\left(0\right)}=\mathcal{F}_{\left(t-1\right),\left\lfloor \left(k-1\right)/\prod_{t'=1}^{t-1}K_{t'}\right\rfloor +1}^{\star}$

$\quad$$\quad$$\:\:$$\:\:$$\:\:$$\:\:$$\:\:$$\:\:$$\:\:$$\:\:$$\:\:$$\:\:$$\:\:$$\:\:$$\:\:$$\:\:$$\:\:$$\:\:$$\mathbf{g}_{\left(t\right),k}^{\left(0\right)}=\mathbf{g}_{\left(t-1\right),\left\lfloor \left(k-1\right)/\prod_{t'=1}^{t-1}K_{t'}\right\rfloor +1}^{\star}$;

$\quad${\scriptsize 5:} $\:\:$$\:\:$$i=0$,

$\quad${\scriptsize 6:} $\:\:$$\:\:$\textbf{repeat}

$\quad$$\quad$$\:\:$$\:\:$$\:\:$$i=i+1$;

$\quad${\scriptsize 7:} $\:\:$$\:\:$$\:\:$Find $\mathbf{g}_{\left(t\right),k}^{\left(i\right)}$
with $\mathcal{F}_{\left(t\right),k}^{\left(i-1\right)}$ by \eqref{eq:problem-1}
to \eqref{eq:Cons1-1};

$\quad${\scriptsize 8:} $\:\:$$\:\:$$\:\:$\textbf{repeat}

$\quad${\scriptsize 9:} $\:\:$$\:\:$$\:\:$$\:\:$\textbf{for} $n=1\colon N$

$\,\,\,\,${\scriptsize 10:} $\:\:$$\:\:$$\:\:$$\:\:$$\:\:$Find
$f_{\left(t\right),n}$, $\forall n$ by \eqref{eq:problem-2-1} to
\eqref{eq:Cons1-2-1};

$\,\,\,\,${\scriptsize 11:} $\:\:$$\:\:$$\:\:$$\:\:$\textbf{end}

$\,\,\,\,${\scriptsize 12:} $\:\:$$\:\:$$\:\:$$\:\:$\textbf{if}
$\mathcal{F}_{\left(t\right),k}=\mathcal{F}_{\left(t\right),k}^{\left(i\right)}$

$\,\,\,\,${\scriptsize 13:} $\:\:$$\:\:$$\:\:$$\:\:$$\:\:$\textbf{break};

$\,\,\,\,${\scriptsize 14:} $\:\:$$\:\:$$\:\:$$\:\:$\textbf{else}

$\,\,\,\,${\scriptsize 15:} $\:\:$$\:\:$$\:\:$$\:\:$$\:\:$$\mathcal{F}_{\left(t\right),k}^{\left(i\right)}=\mathcal{F}_{\left(t\right),k}$;

$\,\,\,\,${\scriptsize 16:} $\:\:$$\:\:$$\:\:$$\:\:$\textbf{end}

$\,\,\,\,${\scriptsize 17:} $\:\:$$\:\:$$\:\:$\textbf{end repeat}

$\,\,\,\,${\scriptsize 18:} $\:\:$$\:\:$\textbf{until} $\mathcal{F}_{\left(t\right),k}^{\left(i\right)}=\mathcal{F}_{\left(t\right),k}^{\left(i-1\right)}$
and $\mathbf{g}_{\left(t\right),k}^{\left(i\right)}=\mathbf{g}_{\left(t\right),k}^{\left(i-1\right)}$;

$\,\,\,\,${\scriptsize 19:} $\:\:$$\:\:$$\mathcal{F}_{\left(t\right),k}^{\star}=\mathcal{F}_{\left(t\right),k}^{\left(i\right)}$
and $\mathbf{g}_{\left(t\right),k}^{\star}=\mathbf{g}_{\left(t\right),k}^{\left(i\right)}$;

$\,\,\,\,${\scriptsize 20:} $\:\:$\textbf{end}

$\,\,\,\,${\scriptsize 21:} \textbf{end}

\textbf{Output: $\mathcal{X}_{k}^{\star}=\left(\mathcal{F}_{k}^{\star},\mathbf{g}_{k}^{\star}\right)=\left(\mathcal{F}_{\left(T\right),k}^{\left(i\right)},\mathbf{g}_{\left(T\right),k}^{\left(i\right)}\right)$},
$k=1,...,K$;
\end{algorithm}
We then update the activated feeding port indices $\mathcal{F}^{\left(i\right)}$
by using $\mathbf{g}^{\left(i\right)}$ obtained from problem \eqref{eq:problem-1}-\eqref{eq:Cons1-1}.
This can be formulated as
\begin{align}
\underset{\mathcal{F}^{\left(i\right)}}{\mathrm{min}}\:\:\mathrm{max}\:\: & \sqrt{\mathrm{Tr}\left(\mathbf{C}_{\mathrm{RPA}}\left(\mathcal{F}^{\left(i\right)},\mathbf{g}^{\left(i\right)},\Omega\right)\right)},\label{eq:problem-2}\\
\mathrm{s.t.}\:\: & \mathrm{card}\left(\mathcal{F}^{\left(i\right)}\right)=N,\label{eq:Cons1-2}
\end{align}
which is an NP-hard optimization problem. To solve this problem, we
use a low-complexity algorithm to sequentially update each entry in
$\mathcal{F}^{\left(i\right)}$. Specifically, we let the optimal
index set at the $\left(i-1\right)$th iteration, $\mathcal{F}^{\left(i-1\right)}$,
to be the initial solution and repeatedly take $n$ as 1 to $N$.
When optimizing the $n$th activated feeding port index, the other
$N-1$ indices of activated feeding ports are fixed. The $n$th activated
feeding port index $f_{n}$ is selected from the remaining $M-N+1$
potential feeding ports indices in $\mathcal{M}$ to minimize the
maximum CRLB of azimuth and elevation angle estimation \eqref{eq:problem-2},
which can be formulated as
\begin{align}
\underset{f_{n}}{\mathrm{min}}\:\:\mathrm{max}\:\: & \sqrt{\mathrm{Tr}\left(\mathbf{C}_{\mathrm{RPA}}\left(\mathcal{F},\mathbf{g}^{\left(i\right)},\Omega\right)\right)},\label{eq:problem-2-1}\\
\mathrm{s.t.}\:\: & f_{n}\in\left(\mathcal{M}\setminus\mathcal{F}^{\left(i\right)}\right)\cup f_{n}^{\left(i\right)},\label{eq:Cons1-2-1}\\
\:\: & \left\{ \mathcal{F}\setminus f_{n}\right\} =\left\{ \mathcal{F}^{\left(i\right)}\setminus f_{n}^{\left(i\right)}\right\} .
\end{align}
The objective function value in \eqref{eq:problem-2-1} is descending
by sequentially updating each entry in $\mathcal{F}$ with optimal
selection to minimize \eqref{eq:problem-2-1}. The algorithm stops
when all entries in activated feeding port indices are no longer changed
and thus we let $\mathcal{F}^{\left(i\right)}=\mathcal{F}$. By iteratively
optimizing the pixel connection states and activated feeding port
positions, the objective function \eqref{eq:problem} to \eqref{eq:Cons2}
can converge to a local optimal solution \cite{Bezdek2003}.

\begin{figure}[t]
\begin{centering}
\includegraphics[width=7.5cm]{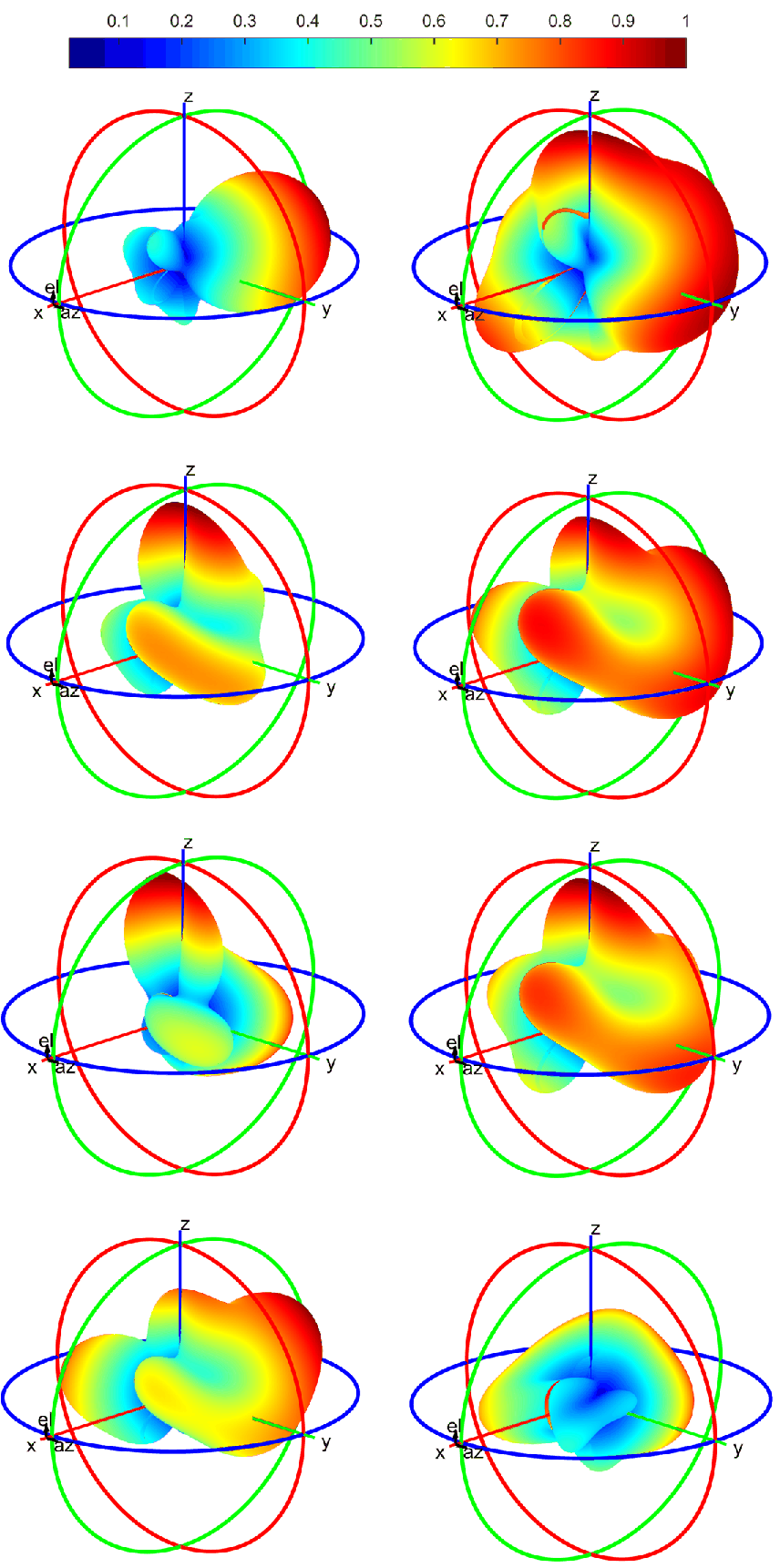}
\par\end{centering}
\caption{Radiation patterns of the proposed HRPA with the optimized geometry
configurations for optimal sensing at broadside angle $\theta=90^{\circ}$
and $\phi=0^{\circ}$ using $N=8$ activated feeding ports. \label{fig:Pattern}}
\end{figure}
\begin{figure*}[t]
\begin{centering}
\includegraphics[width=8cm]{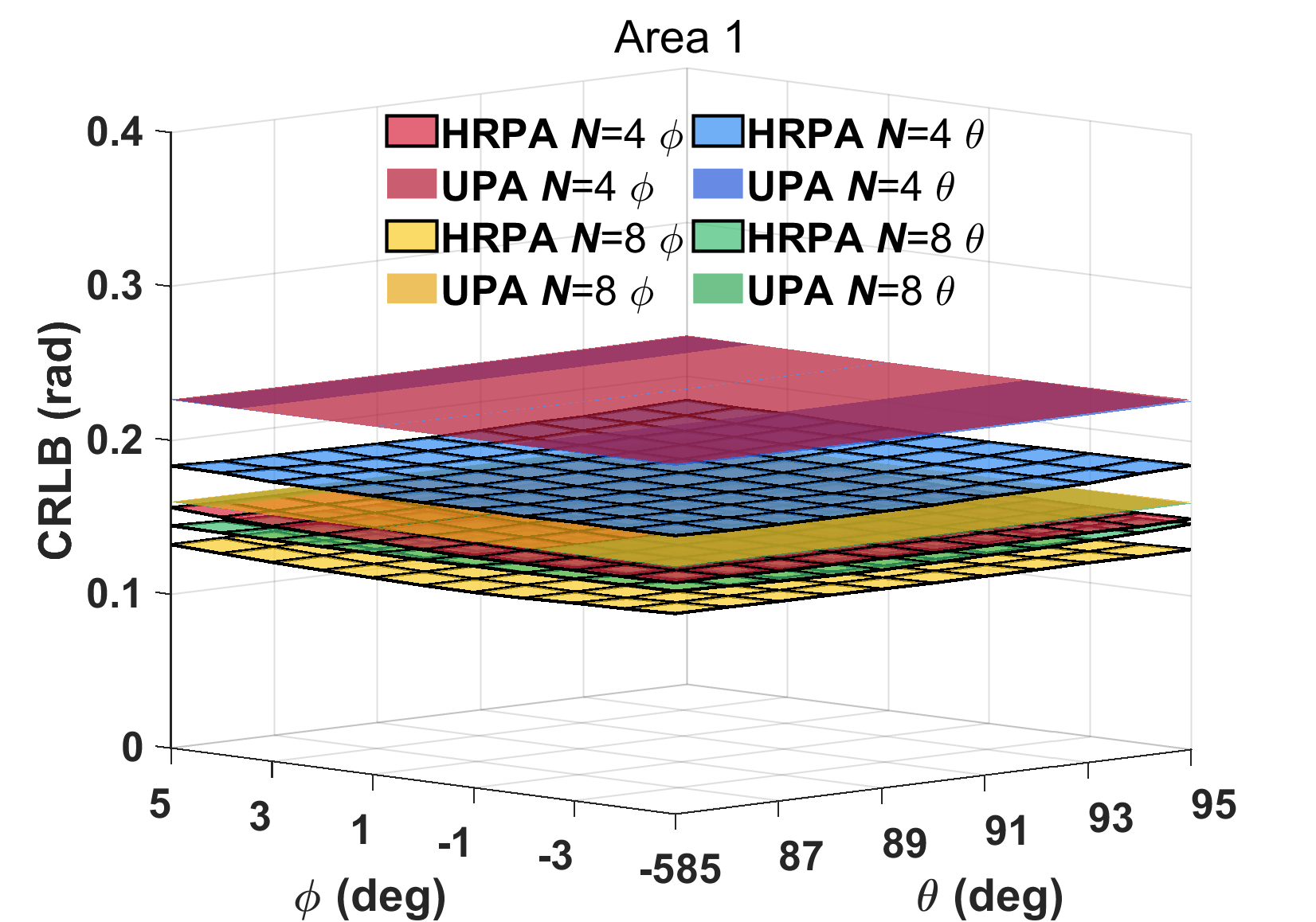}\includegraphics[width=8cm]{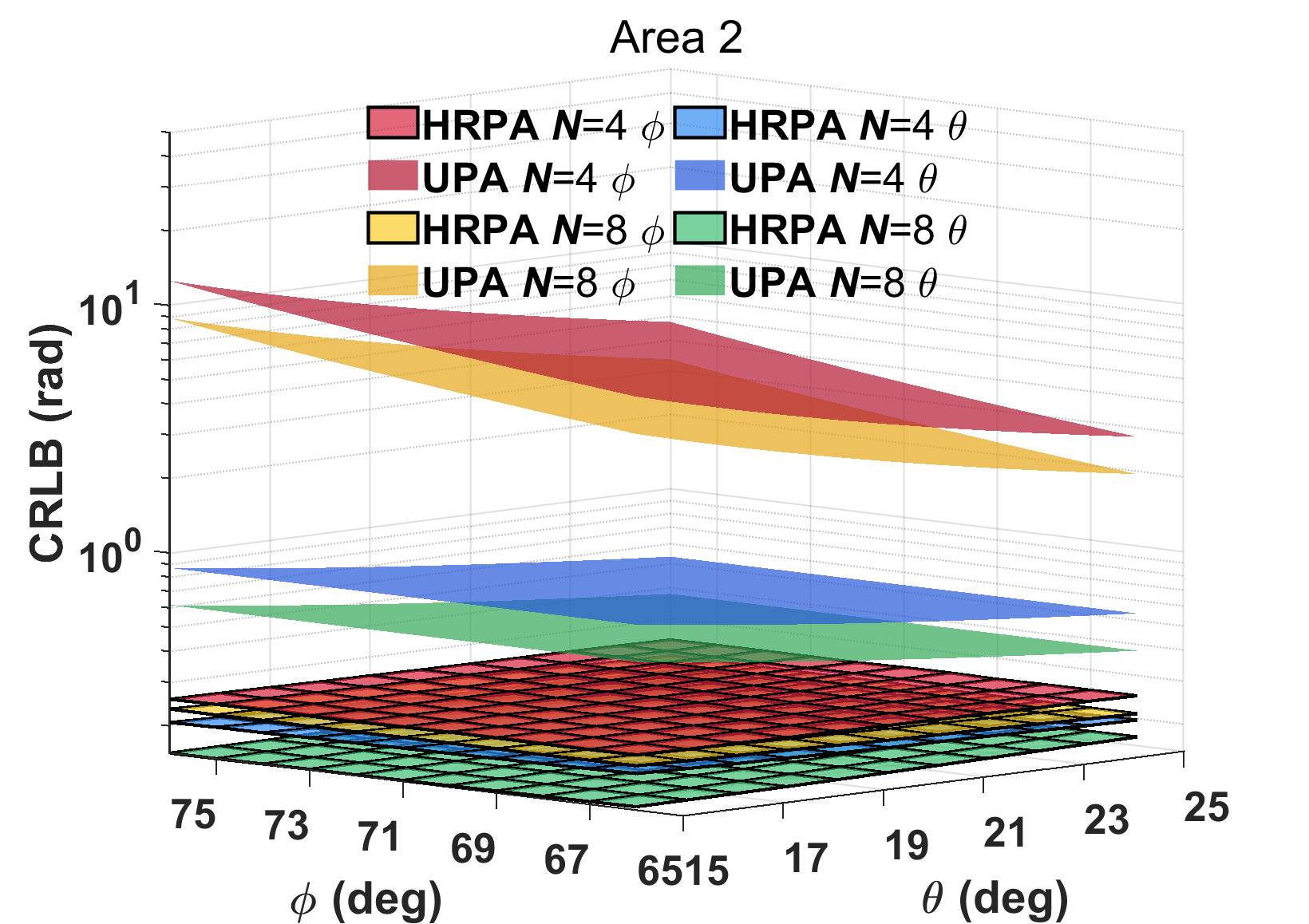}
\par\end{centering}
\begin{centering}
\includegraphics[width=8cm]{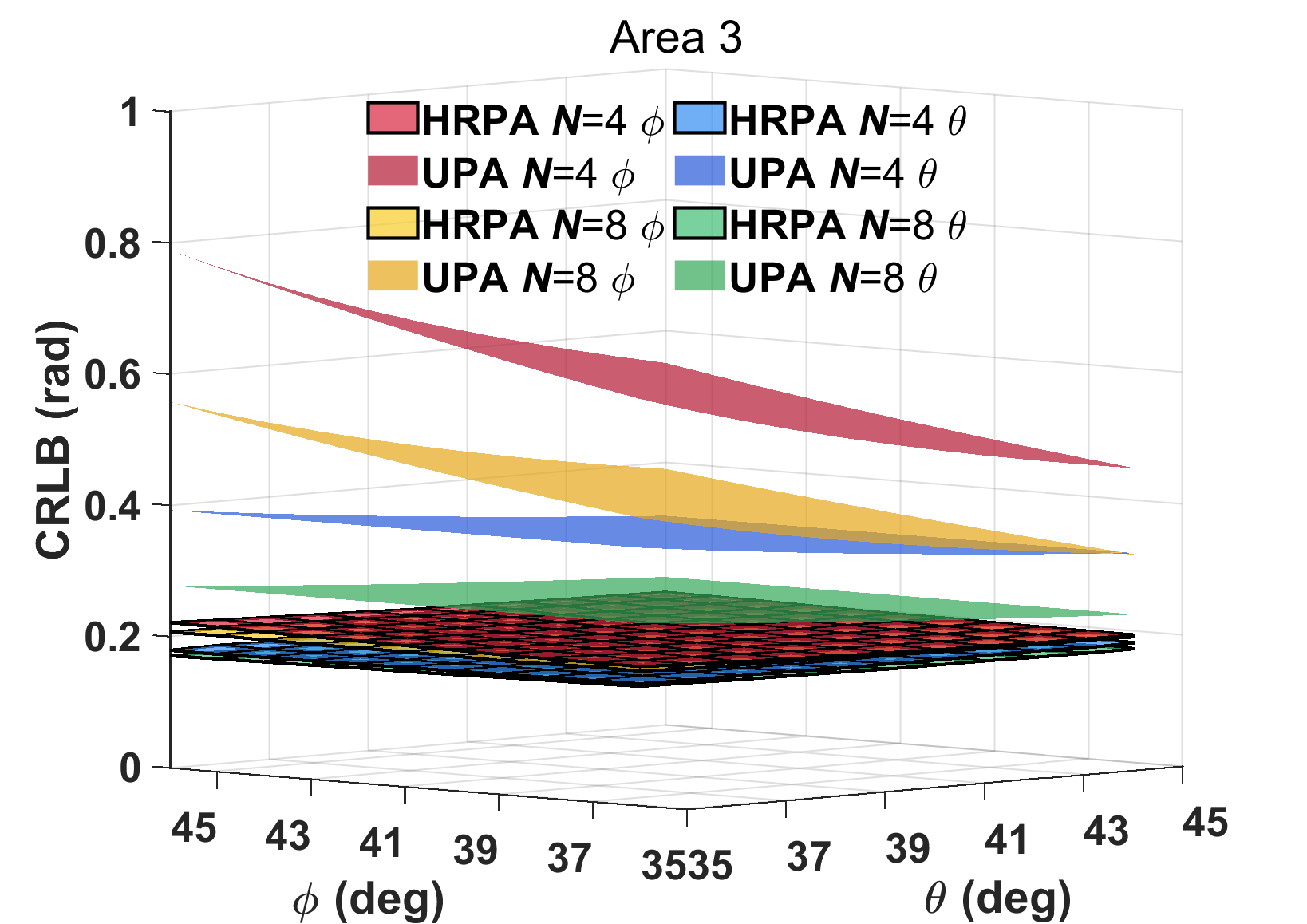}\includegraphics[width=8cm]{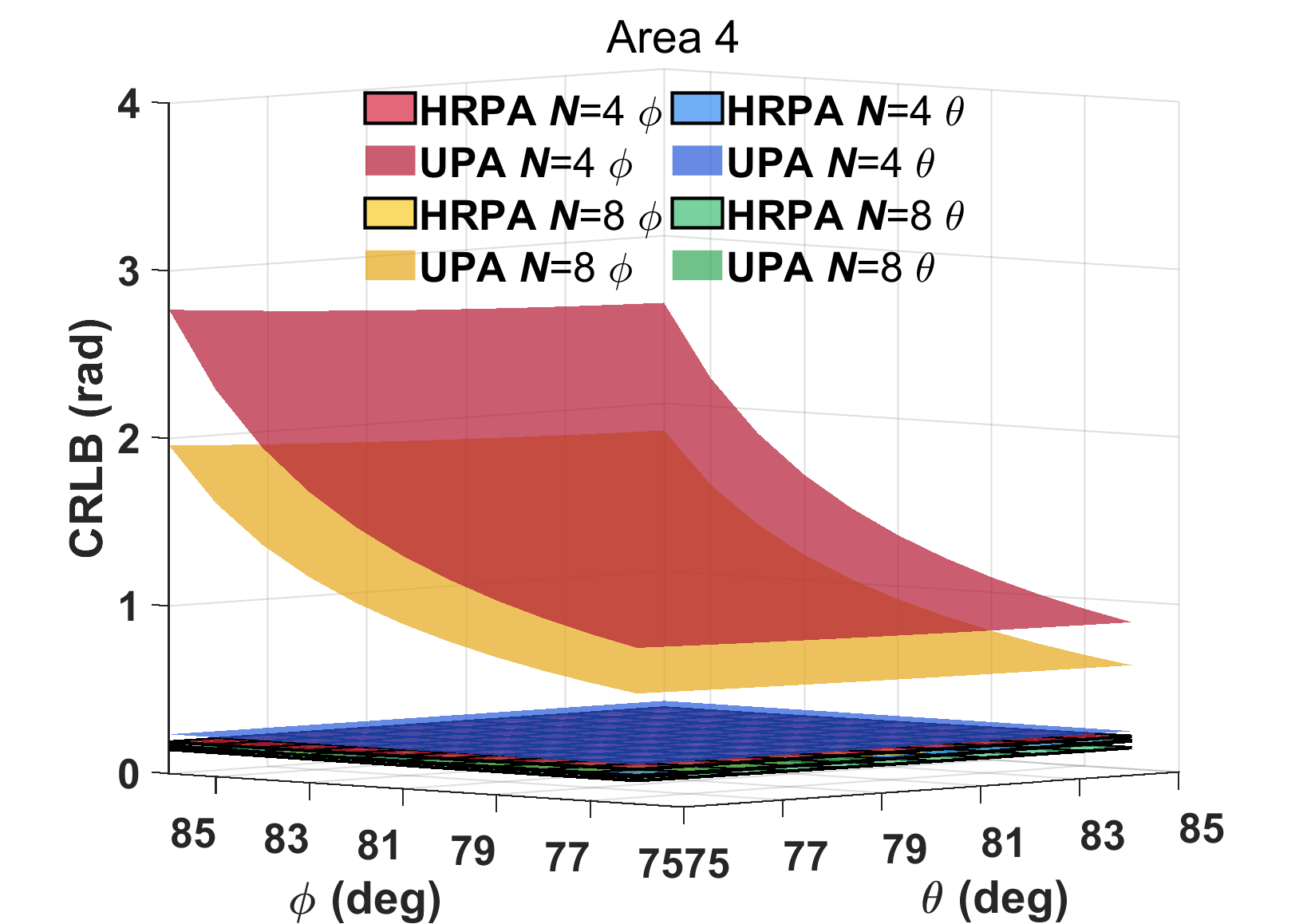}
\par\end{centering}
\caption{CRLB of HRPA with $N=4$ and 8 activated feeding ports in four different
sensing areas. \label{fig:Pattern vs UPA}}
\end{figure*}
In the practical implementation, we wish to divide the total radiation
space $\Omega$ into $K$ sensing areas $\Omega_{k}$, $k=1,2,...,K$
which satisfy $\bigcup_{k=1}^{K}\Omega_{k}=\Omega$ and $\Omega_{k}\bigcap\Omega_{l}=\emptyset$
for $k\neq l$. When the sensing target locates in the area $\Omega_{k}$,
the HRPA can use corresponding geometry configuration to excite $N$
radiation patterns for achieving optimal angular sensing accuracy.
To that end, we need to find $K$ codewords $\mathcal{X}_{k}=\left(\mathcal{F}_{k},\mathbf{g}_{k}\right)$,
$k=1,2,...,K$, as the optimal HRPA geometry configurations for $K$
sensing area $\Omega_{k}$. To efficiently find $K$ codewords, we
propose to use a subdivision method. Specifically at the initial stage,
we start from optimizing the HRPA geometry configuration, using the
proposed alternating optimization method and initial guess $\mathcal{X}_{\left(0\right)}^{\star}=\left(\mathcal{F}_{\left(0\right)}^{\star},\mathbf{g}_{\left(0\right)}^{\star}\right)$,
for the whole radiation space $\Omega_{\left(1\right)}=\bigcup_{k=1}^{K}\Omega_{k}$
where the optimal codeword is denoted as $\mathcal{X}_{\left(1\right)}^{\star}$.
Then at the second stage, we can divide the space $\Omega_{\left(1\right)}$
into $K_{2}$ areas $\Omega_{\left(2\right),k}$, $k=1,2,...,K_{2}$
with the same size satisfying $\bigcup_{k=1}^{K_{2}}\Omega_{\left(2\right),k}=\Omega$
and $\Omega_{\left(2\right),k}\bigcap\Omega_{\left(2\right),l}=\emptyset$
for $k\neq l$, e.g., when $K_{2}=2$ we have $\Omega_{\left(2\right),1}=\bigcup_{k=1}^{K/2}\Omega_{k}$
and $\Omega_{\left(2\right),2}=\bigcup_{k=K/2+1}^{K/2}\Omega_{k}$.
To optimize the codewords for each of the $K_{2}$ areas, we use the
$\mathcal{X}_{\left(1\right)}^{\star}$ as the initial guess in the
alternating optimization method and obtain the optimal codewords as
$\mathcal{X}_{\left(2\right),k}^{\star}$, $k=1,2,...,K_{2}$, which
can be again used as the initial guess for the next subdivision stage.
By repeating this process to divide each of the area $\Omega_{\left(t\right),k}$
into $K_{t+1}$ areas at the $t+1$ stages, we have totally $K$ areas
for $T$ stages $\prod_{t=1}^{T}K_{t}=K$ ($K_{1}$ is set as unity).
The $K$ codewords of the optimal HRPA geometry configurations, $\mathcal{F}_{k}^{\star}$
and $\mathbf{g}_{k}^{\star}$, $k=1,2,...,K$, for $K$ areas form
a codebook that can be used in angular sensing.

Algorithm \ref{alg:AO} summarizes the overall algorithm for optimizing
the activated feeding port indices and pixel connection states. The
performance of the algorithm in finding optimal HRPA geometries will
be shown in the next section.

\section{Numerical Results}

\subsection{HRPA Design and Optimization}

To demonstrate the performance of the proposed HRPA in angular sensing,
we utilize the analysis and optimization approach in Section III and
IV to obtain the optimal geometries of HRPA. The HRPA architecture
in Fig. \ref{fig:Pixel Antenna} is used, which can be viewed as discritizing
the metal surface of a conventional patch antenna into a $5\times5$
pixel array. The side length of substrate and pixels are $w_{1}=62.5$
mm and $w_{2}=12$ mm respectively. The height of the HRPA is $h=12.5$
mm. The HRPA has totally 40 loaded ports that can be implemented by
RF switches to connect adjacent pixels and 25 potential feeding ports
across the pixel array and copper ground. The material of substrate
is Rogers 4003C which has thickness of 1.524 mm, permittivity of 3.55
and loss tangent of 0.0027 while the metal is made of copper which
has electric conductivity of $5.8\times10^{7}$ S/m.

We then utilize the Algorithm 1 to obtain the optimal geometries of
the HRPA for exciting $N$ radiation patterns. Specifically, the population
size and generation number of GA are set as 500 and 200, respectively.
SNR is a constant coefficient in \eqref{eq:CRB} which is set as 0
dB in the following simulation. The resolution of both azimuth and
elevation angle are set as $1^{\circ}$. The inner impedance of sources
connected to the activated feeding ports, i.e., diagonal entries in
$\mathbf{Z}_{0}$, are set as 50 Ohm. 

In Fig. \ref{fig:Pattern}, we provide the radiation patterns of the
proposed HRPA with the optimized geometries for optimal sensing at
broadside angle $\theta=90^{\circ}$ and $\phi=0^{\circ}$ using $N=8$
activated feeding ports for reference. These are plotted in the full
3D sphere in the far field. We can notice that these radiation patterns
are quite different at broadside direction, making the angular gradient
large at the broadside angle. Radiation patterns for the other numbers
of activated feeding ports $N$ and sensing areas follow a similar
trend so we omit them.

\subsection{Performance of HRPA in Angular Sensing}

To demonstrate the proposed HRPA in angular sensing, we utilize the
radiation patterns, as illustrated in Fig. \ref{fig:Pattern}, to
evaluate the performance of HRPA as the MIMO sensing receiver. 

\subsubsection{CRLB of HRPA in Different Sensing Areas}

We firstly consider the HRPA with $N=4$ and 8 activated feeding ports.
They are benchmarked with a conventional $2\times2$ half-wavelength
spaced UPA with single and dual polarization. In addition, we set
the size of each sensing area as $10^{\circ}\times10^{\circ}$, that
is, both azimuth and elevation angle range of each area are $10^{\circ}$.
The CRLB results of optimized HRPA geometries, with unit of radians,
in angular sensing are shown in Fig. \ref{fig:Pattern vs UPA} where
four sensing areas in different regions of 3D sphere are illustrated,
including 1) a broadside angle range $\theta\in\left[85^{\circ},95^{\circ}\right]$
and $\phi\in\left[-5^{\circ},5^{\circ}\right]$, three endfire angle
ranges 2) $\theta\in\left[15^{\circ},25^{\circ}\right]$ and $\phi\in\left[65^{\circ},75^{\circ}\right]$,
3) $\theta\in\left[35^{\circ},45^{\circ}\right]$ and $\phi\in\left[35^{\circ},45^{\circ}\right]$,
and 4) $\theta\in\left[75^{\circ},85^{\circ}\right]$ and $\phi\in\left[75^{\circ},85^{\circ}\right]$.
We can make the following observations.

\textit{Firstly}, it can be straightforwardly observed that by using
the HRPA, the angular sensing accuracy of MIMO receivers for both
azimuth and elevation angles can be significantly enhanced in all
sensing areas when compared to conventional UPA with fixed antenna
geometry. The essence of this performance enhancement is that the
geometries of HRPA can be optimized to maximize the pattern angular
gradient of HRPA at the given angle, resulting in lower CRLBs. In
particular, the angle estimation accuracy can be improved by around
50\% in the four areas for both elevation or azimuth angle, demonstrating
the effectiveness of the HRPA on enhancing angular sensing of MIMO
system.

\textit{Secondly}, we can observe that the CRLB of HRPA in four sensing
areas are generally around 0.2 radians while CRLBs of conventional
UPA drastically change in various areas, demonstrating that the HRPA
has stable sensing capability for different directions. However, the
performance of UPA is more sensitive to the AoA of the sensing targets,
i.e., UPA performs best at the broadside angle while its performance
rapidly degrades at the endfire angles. For this reason, the performance
improvement by the HRPA at the endfire angles is the most significant.

\textit{Thirdly}, by comparing the results of $N=4$ and $N=8$, we
can find that the improvement of CRLB for the HRPA from $N=4$ to
$8$ is below 50\% in all areas while CRLB of conventional UPA can
be halved by using double number of feeding ports. This is because
mutual coupling in HRPA becomes stronger when using more activated
feeding ports so that radiation efficiencies are reduced, affecting
the consequent CRLBs of angle estimation.

\subsubsection{CRLB of HRPA with Different Area Size}

\begin{figure}[t]
\begin{centering}
\includegraphics[width=8cm]{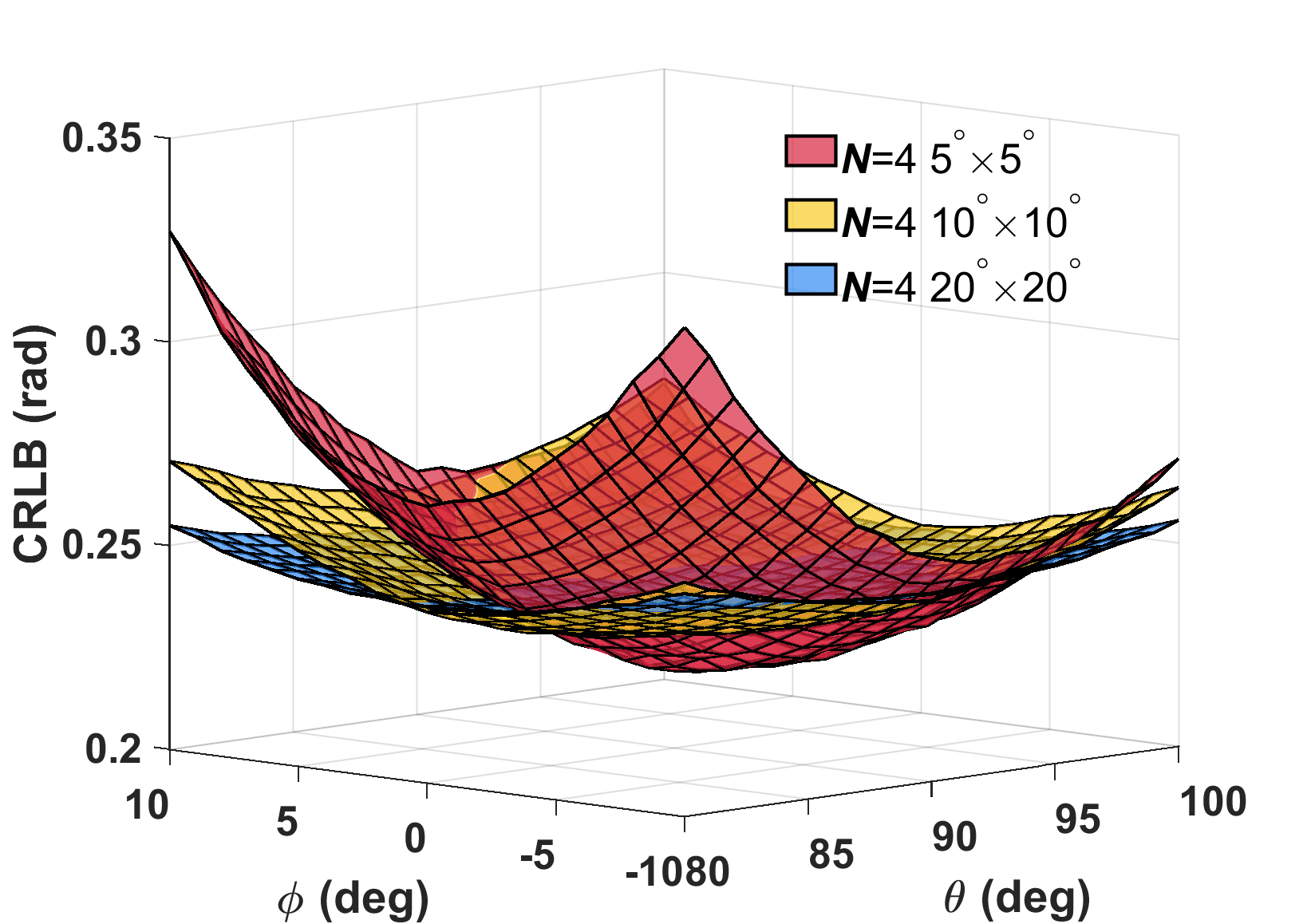}
\par\end{centering}
\begin{centering}
(a)
\par\end{centering}
\begin{centering}
\includegraphics[width=8cm]{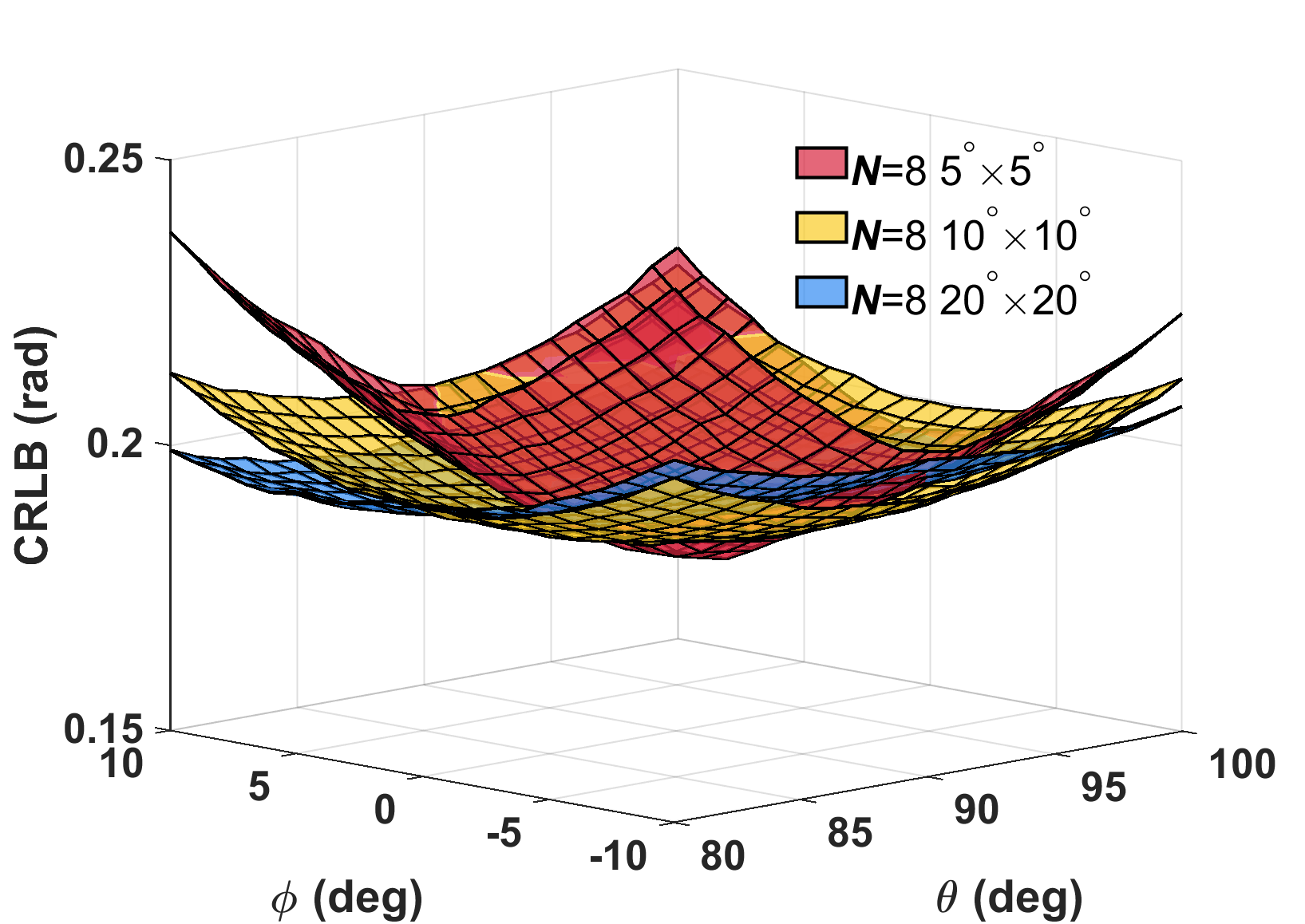}
\par\end{centering}
\begin{centering}
(b)
\par\end{centering}
\caption{CRLBs of HRPA with (a) $N=4$ and (b) $N=8$ activated feeding ports
when different size of sensing areas in the optimization are used.
\label{fig:CRLB size}}
\end{figure}
Considering that when the target moves to different sensing areas,
HRPA needs to switch its geometry configurations to track the target.
An HRPA geometry that can cover a larger area while maintaining sensing
performance is desired since the switching speed of HRPA geometry
configuration can be slower. Therefore, we also investigate the effect
of different area size on CRLB of the optimal HRPA geometries. Specifically
for cases of $N=4$ and 8 activated feeding ports, we select three
area size of $5^{\circ}\times5^{\circ}$, $10^{\circ}\times10^{\circ}$
and $20^{\circ}\times20^{\circ}$ in the optimization. These three
areas are centered at the broadside direction $\theta=90^{\circ}$
and $\phi=0^{\circ}$. Therefore in the optimization algorithm, the
area with the largest size, $20^{\circ}\times20^{\circ}$, will be
firstly optimized and its optimal geometry configuration is used as
the initial point of the optimization afterwards for areas with smaller
size $10^{\circ}\times10^{\circ}$ and $5^{\circ}\times5^{\circ}$.

We compare the CRLB results of three geometries, in terms of the overall
solid angle estimation (equals to the objective function value in
\eqref{eq:problem}), in the same angle range of $\theta\in\left[80^{\circ},90^{\circ}\right]$
and $\phi\in\left[-10^{\circ},10^{\circ}\right]$ as shown in Fig.
\ref{fig:CRLB size}. Two observations can be made.

\textit{Firstly}, it can be observed that from $20^{\circ}\times20^{\circ}$
case to $5^{\circ}\times5^{\circ}$ case, CRLBs of the HRPA are improved
within the optimized sensing area, showing that the angular sensing
accuracy can be improved when a smaller area size is used in the optimization.
However, the number of codewords $K$ will be larger when with smaller
area size so that more HRPA geometry configurations need to be optimized
and switching the HRPA geometry is more frequent.

\textit{Secondly}, we can observe that the CRLB performance of the
HRPA degrades when outside the sensing area in the optimization, indicating
that the performance enhancement by HRPA within the sensing area is
achieved at the expense of the performance degradation in the other
areas. A $10^{\circ}$ range of azimuth and elevation angles performs
good tradeoff between the area size and sensing performance.

\subsubsection{CRLB Tradeoff of HRPA}

\begin{figure}[t]
\begin{centering}
\includegraphics[width=8.5cm]{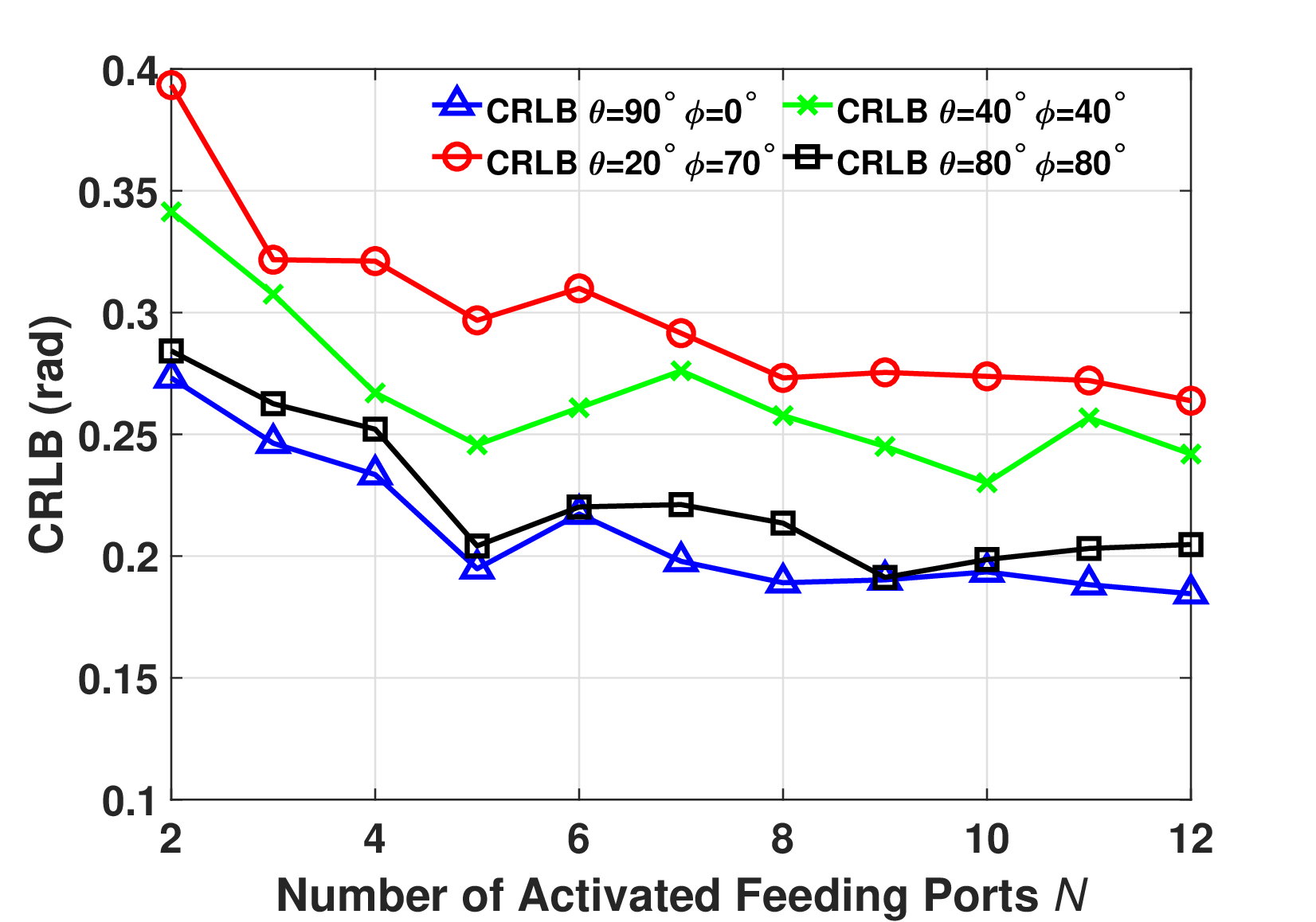}
\par\end{centering}
\caption{CRLBs of the HRPA when different number of activated feeding ports
are used. \label{fig:Tradeoff}}
\end{figure}

To further investigate the relationship between the number of activated
feeding ports $N$ and CRLB of the optimal HRPA geometries, we provide
CRLB of the overall solid angle estimation, as shown in Fig. \ref{fig:Tradeoff},
when different number of activated feeding ports $N$ are used. We
can make the following observations.

\textit{Firstly}, we can observe that when $N<8$, CRLBs of the HRPA
at different angles decreases as the number of activated feeding ports
$N$ increases. This is because using more activated feeding ports
enables a denser spatial sampling through more radiation patterns
and thus the pattern angular gradient can be enhanced to reduce CRLB.

\textit{Secondly}, we can observe that CRLBs approach convergence
when $N\geqslant8$. The reason is that the efficiencies of partial
activated feeding ports are very low so that the radiation patterns
of these ports make minor contributions to the angular sensing. Therefore,
for the designed HRPA with side length of half wavelength, $N=8$
activated feeding ports can perform a good tradeoff between the angular
sensing accuracy and implementation complexity. 

\textit{Thirdly}, it can be observed that CRLBs of various angle directions
are generally close for all number of activated feeding ports, demonstrating
again that the proposed HRPA has stable sensing capability for arbitrary
angle directions.

To conclude, by using the HRPA to perform angular sensing, the angle
estimation accuracy of MIMO receivers can be enhanced. The key advantage
of the proposed HRPA approach is that the optimal geometry configurations
can be adjusted for targets in various sensing areas to enhance angle
estimation performance. The analysis above can provide a guidance
for HRPA design and optimization in angular sensing.

\section{Conclusions}

In this paper, we propose a novel HRPA architecture and investigate
its functionality in angular sensing. The proposed multi-port HRPA
is extended from the conventional pixel antenna by combining advantages
of pixel antennas with extreme radiating aperture reconfigurability
and FASs with highly movable port positions. As a result, the geometries
of the HRPA can be flexibly changed to adjust the radiation patterns
for sensing the angles of targets.

Specifically in this work, the architecture and the equivalent circuit
model of HRPA are provided. The radiation patterns of the multi-port
HRPA are then derived as a function of pixel connections and port
selections. We also develop an efficient alternating optimization
approach to obtain the optimal geometries of the HRPA by minimizing
CRLB. The resulting geometry configurations form a codebook for angular
sensing in different sensing areas. Simulation results show that the
proposed HRPA outperforms conventional UPA with same size over the
full 3D sphere in angular sensing where the angle estimation accuracy
can be improved by more than 50\%. The performance improvement at
the endfire angles is the most significant. These results demonstrate\textcolor{red}{{}
}the effectiveness of the proposed approach and show the promise of
HRPA in the sixth generation (6G) ISAC system.

Based on the proposed model in this paper, the artificial intelligence
(AI) technique can be further utilized to enhance the HRPA optimization
and function in future work. AI is able to perform real-time control
of the pixel connections and port selection of HRPA instead of using
a codebook optimized in an offline process. In addition, AI can learn
the movement of targets and accurately predict their future directions,
thereby reducing the reconfiguration latency in target tracking to
support high-mobility scenarios. For these reasons, AI is a powerful
tool for real-time sensing and tracking and this direction is left
as our first future work.

Another direction for future work is to fabricate a prototype of the
HRPA incorporating both variable loads and port selection. The models
of practical RF switches and feeding ports can be considered in the
optimization and design process. Experimental validation can then
be carried out to evaluate the angular sensing accuracy in real-world
environments. Overall, the general approach proposed in this work
provides a versatile framework for developing compact HRPA designs
across various application scenarios, including mobile devices and
Internet of Things (IoT) systems.

\bibliographystyle{IEEEtran}

\begin{thebibliography}{10}
	\providecommand{\url}[1]{#1}
	\csname url@samestyle\endcsname
	\providecommand{\newblock}{\relax}
	\providecommand{\bibinfo}[2]{#2}
	\providecommand{\BIBentrySTDinterwordspacing}{\spaceskip=0pt\relax}
	\providecommand{\BIBentryALTinterwordstretchfactor}{4}
	\providecommand{\BIBentryALTinterwordspacing}{\spaceskip=\fontdimen2\font plus
		\BIBentryALTinterwordstretchfactor\fontdimen3\font minus
		\fontdimen4\font\relax}
	\providecommand{\BIBforeignlanguage}[2]{{%
			\expandafter\ifx\csname l@#1\endcsname\relax
			\typeout{** WARNING: IEEEtran.bst: No hyphenation pattern has been}%
			\typeout{** loaded for the language `#1'. Using the pattern for}%
			\typeout{** the default language instead.}%
			\else
			\language=\csname l@#1\endcsname
			\fi
			#2}}
	\providecommand{\BIBdecl}{\relax}
	\BIBdecl
	
	\bibitem{Zhang2019}
	Z.~Zhang, Y.~Xiao, Z.~Ma, M.~Xiao, Z.~Ding, X.~Lei, G.~K. Karagiannidis, and
	P.~Fan, ``6{G} wireless networks: Vision, requirements, architecture, and key
	technologies,'' \emph{IEEE Veh. Technol. Mag.}, vol.~14, no.~3, pp. 28--41,
	2019.
	
	\bibitem{10530985}
	W.~Xu, C.~Zhao, and F.~Gao, ``Angle domain channel-based camera pose correction
	for vision-aided {ISAC} systems,'' \emph{IEEE Wirel. Commun. Lett.}, vol.~13,
	no.~8, pp. 2080--2084, 2024.
	
	\bibitem{Han2023}
	Z.~Han, H.~Ding, X.~Zhang, Y.~Wang, M.~Lou, J.~Jin, Q.~Wang, and G.~Liu,
	``Multistatic integrated sensing and communication system in cellular
	networks,'' in \emph{2023 IEEE Globecom Workshops (GC Wkshps)}, 2023, pp.
	123--128.
	
	\bibitem{1367557}
	B.~Cetiner, H.~Jafarkhani, J.-Y. Qian, H.~J. Yoo, A.~Grau, and F.~De~Flaviis,
	``Multifunctional reconfigurable {MEMS} integrated antennas for adaptive
	{MIMO} systems,'' \emph{IEEE Commun. Mag.}, vol.~42, no.~12, pp. 62--70,
	2004.
	
	\bibitem{Tang2021}
	S.~Tang, Y.~Zhang, Z.~Han, C.-Y. Chiu, and R.~Murch, ``A pattern-reconfigurable
	antenna for single-{RF} 5{G} millimeter-wave communications,'' \emph{IEEE
		Antennas Wirel. Propag. Lett.}, vol.~20, no.~12, pp. 2344--2348, 2021.
	
	\bibitem{Zhang2022}
	Y.~Zhang, Z.~Han, S.~Tang, S.~Shen, C.-Y. Chiu, and R.~Murch, ``A highly
	pattern-reconfigurable planar antenna with 360 single- and multi-beam
	steering,'' \emph{IEEE Trans. Antennas Propag.}, vol.~70, no.~8, pp.
	6490--6504, 2022.
	
	\bibitem{Tang2023}
	S.~Tang, Y.~Zhang, J.~Rao, Z.~Han, C.-Y. Chiu, and R.~Murch, ``Beamforming
	network design utilizing node microstrip architectures for dual-polarized
	endfire millimeter-wave antenna arrays,'' \emph{IEEE Trans. Antennas
		Propag.}, vol.~71, no.~6, pp. 4862--4873, 2023.
	
	\bibitem{Zhang2022a}
	Y.~Zhang, S.~Tang, Z.~Han, J.~Rao, S.~Shen, M.~Li, C.-Y. Chiu, and R.~Murch,
	``A low-profile microstrip vertically polarized endfire antenna with 360
	beam-scanning and high beam-shaping capability,'' \emph{IEEE Trans. Antennas
		Propag.}, vol.~70, no.~9, pp. 7691--7702, 2022.
	
	\bibitem{Jing2022}
	L.~Jing, M.~Li, and R.~Murch, ``Compact pattern reconfigurable pixel antenna
	with diagonal pixel connections,'' \emph{IEEE Trans. Antennas Propag.},
	vol.~70, no.~10, pp. 8951--8961, 2022.
	
	\bibitem{10669211}
	Y.~Zhang, S.~Tang, J.~Rao, C.-Y. Chiu, X.~Chen, and R.~Murch, ``A dual-port
	dual-beam pattern-reconfigurable antenna with independent 2-d
	beam-scanning,'' \emph{IEEE Transactions on Antennas and Propagation},
	vol.~72, no.~10, pp. 7628--7643, 2024.
	
	\bibitem{Rao2022}
	J.~Rao, Y.~Zhang, S.~Tang, Z.~Li, S.~Shen, C.-Y. Chiu, and R.~Murch, ``A novel
	reconfigurable intelligent surface for wide-angle passive beamforming,''
	\emph{IEEE Trans. Microwave Theory Tech.}, vol.~70, no.~12, pp. 5427--5439,
	2022.
	
	\bibitem{Rao2023}
	J.~Rao, Y.~Zhang, S.~Tang, Z.~Li, C.-Y. Chiu, and R.~Murch, ``An active
	reconfigurable intelligent surface utilizing phase-reconfigurable reflection
	amplifiers,'' \emph{IEEE Trans. Microwave Theory Tech.}, vol.~71, no.~7, pp.
	3189--3202, 2023.
	
	\bibitem{Zhang2021}
	Y.~Zhang, S.~Shen, Z.~Han, C.-Y. Chiu, and R.~Murch, ``Compact {MIMO} systems
	utilizing a pixelated surface: Capacity maximization,'' \emph{IEEE Trans.
		Veh. Technol.}, vol.~70, no.~9, pp. 8453--8467, 2021.
	
	\bibitem{Zhang2020}
	Y.~Zhang, Z.~Han, S.~Shen, C.-Y. Chiu, and R.~Murch, ``Polarization enhancement
	of microstrip antennas by asymmetric and symmetric grid defected ground
	structures,'' \emph{IEEE Open J. Antennas Propag.}, vol.~1, pp. 215--223,
	2020.
	
	\bibitem{Shen2024}
	S.~Shen, K.-K. Wong, and R.~Murch, ``Antenna coding empowered by pixel
	antennas,'' \emph{IEEE Trans. Commun.}, vol.~74, pp. 446--460, 2026.
	
	\bibitem{han2025exploiting}
	Z.~Han, S.~Shen, and R.~Murch, ``Exploiting spatial multiplexing based on pixel
	antennas: An antenna coding approach,'' \emph{arXiv preprint
		arXiv:2512.05706}, 2025.
	
	\bibitem{9264694}
	K.-K. Wong, A.~Shojaeifard, K.-F. Tong, and Y.~Zhang, ``Fluid antenna
	systems,'' \emph{IEEE Trans. Wirel. Commun.}, vol.~20, no.~3, pp. 1950--1962,
	2021.
	
	\bibitem{Yang2024}
	H.~Yang, H.~Xu, K.-K. Wong, C.-B. Chae, R.~Murch, S.~Jin, and Y.~Zhang,
	``Position index modulation for fluid antenna system,'' \emph{IEEE Trans.
		Wirel. Commun.}, vol.~23, no.~11, pp. 16\,773--16\,787, 2024.
	
	\bibitem{9388928}
	Y.~Huang, L.~Xing, C.~Song, S.~Wang, and F.~Elhouni, ``Liquid antennas: Past,
	present and future,'' \emph{IEEE Open J. Antennas Propag.}, vol.~2, pp.
	473--487, 2021.
	
	\bibitem{10740058}
	J.~Zhang, J.~Rao, Z.~Li, Z.~Ming, C.-Y. Chiu, K.-K. Wong, K.-F. Tong, and
	R.~Murch, ``A novel pixel-based reconfigurable antenna applied in fluid
	antenna systems with high switching speed,'' \emph{IEEE Open J. Antennas
		Propag.}, vol.~6, no.~1, pp. 212--228, 2025.
	
	\bibitem{New2024}
	W.~K. New, K.-K. Wong, H.~Xu, K.-F. Tong, and C.-B. Chae, ``Fluid antenna
	system: New insights on outage probability and diversity gain,'' \emph{IEEE
		Trans. Wirel. Commun.}, vol.~23, no.~1, pp. 128--140, 2024.
	
	\bibitem{New2024a}
	------, ``An information-theoretic characterization of {MIMO-FAS}:
	Optimization, diversity-multiplexing tradeoff and q-outage capacity,''
	\emph{IEEE Trans. Wirel. Commun.}, vol.~23, no.~6, pp. 5541--5556, 2024.
	
	\bibitem{Ma2024}
	W.~Ma, L.~Zhu, and R.~Zhang, ``{MIMO} capacity characterization for movable
	antenna systems,'' \emph{IEEE Trans. Wirel. Commun.}, vol.~23, no.~4, pp.
	3392--3407, 2024.
	
	\bibitem{Ye2024}
	Y.~Ye, L.~You, J.~Wang, H.~Xu, K.-K. Wong, and X.~Gao, ``Fluid antenna-assisted
	{MIMO} transmission exploiting statistical {CSI},'' \emph{IEEE Commun.
		Lett.}, vol.~28, no.~1, pp. 223--227, 2024.
	
	\bibitem{10705114}
	J.~Zou, H.~Xu, C.~Wang, L.~Xu, S.~Sun, K.~Meng, C.~Masouros, and K.-K. Wong,
	``Shifting the {ISAC} trade-off with fluid antenna systems,'' \emph{IEEE
		Wirel.s Commun. Lett.}, vol.~13, no.~12, pp. 3479--3483, 2024.
	
	\bibitem{10707252}
	L.~Zhou, J.~Yao, M.~Jin, T.~Wu, and K.-K. Wong, ``Fluid antenna-assisted {ISAC}
	systems,'' \emph{IEEE Wirel. Commun. Lett.}, vol.~13, no.~12, pp. 3533--3537,
	2024.
	
	\bibitem{1561599}
	H.~Gazzah and S.~Marcos, ``Cramer-{R}ao bounds for antenna array design,''
	\emph{IEEE Trans. Signal Process.}, vol.~54, no.~1, pp. 336--345, 2006.
	
	\bibitem{CMS}
	{CST Microwave Studio 2019}, \emph{http://www.cst.com}.
	
	\bibitem{7762757}
	S.~Shen, Y.~Sun, S.~Song, D.~P. Palomar, and R.~D. Murch, ``Successive boolean
	optimization of planar pixel antennas,'' \emph{IEEE Trans. Antennas Propag.},
	vol.~65, no.~2, pp. 920--925, 2017.
	
	\bibitem{9491941}
	F.~Jiang, S.~Shen, C.-Y. Chiu, Z.~Zhang, Y.~Zhang, Q.~S. Cheng, and R.~Murch,
	``Pixel antenna optimization based on perturbation sensitivity analysis,''
	\emph{IEEE Trans. Antennas Propag.}, vol.~70, no.~1, pp. 472--486, 2022.
	
	\bibitem{10035928}
	T.~Qiao, F.~Jiang, S.~Shen, Z.~Zhang, M.~Li, C.-Y. Chiu, Q.~S. Cheng, and
	R.~Murch, ``Pixel antenna optimization using the adjoint method and the
	method of moving asymptote,'' \emph{IEEE Trans. Antennas Propag.}, vol.~71,
	no.~3, pp. 2873--2878, 2023.
	
	\bibitem{Bezdek2003}
	J.~C. Bezdek and R.~J. Hathaway, ``Convergence of alternating optimization,''
	\emph{Neural, Parallel Sci. Comput.}, vol.~11, no.~4, pp. 351--368, 2003.
	
\end{thebibliography}

\end{document}